\documentclass[hyper]{article}
\usepackage{jheppub}
 \usepackage{enumerate}
 \usepackage{url}
\input{epsf}
\usepackage{epsfig}
\usepackage[all]{xy}
\usepackage{amsmath,amsthm}
\def\be{ \begin{equation} }
\def\ee{ \end{equation}}

\def\Aut{{\rm Aut}}

\def\Co0{{\rm Co}_0}

\def\dim{{\rm dim}}
\def\End{{\rm End}}
\def\exp{{\rm exp}}

\def\I{{\rm i}}

\renewcommand{\Im}{{\rm Im }}

\def\mod{{\rm mod}}
\def\Pin{{\rm Pin}}

\def\Tr{{\rm Tr}}

\def\half{\frac{1}{2}}
\def\p{\partial}

\def\one{{\hbox{ 1\kern-.8mm l}}}

\def\CC {{\cal C}}
\def\CD {{\cal D}}
\def\CE {{\cal E}}

\def\CG {{\cal G}}
\def\CH {{\cal H}}

\def\CO {{\cal O}}

\def\CV {{\cal V}}

\def\CO {{\cal O}}

\def\CE {{\cal E}}
\def\CG {{\cal G}}
\def\CH {{\cal H}}

\def\CS {{\cal S}}

\def\IC{\mathbb{C}}

\def\IF{\mathbb{F}}

\def\IH{\mathbb{H}}

\def\IR{{\mathbb{R}}}

\def\IZ{{\mathbb{Z}}}

\def\ff{\mathfrak{f}}

\def\fo{\mathfrak{o}}

\def\fs{\mathfrak{s}}

\def\fs{\mathfrak{s}}

\def\fu{\mathfrak{u}}

\def\rmk#1{\bigskip\noindent{\bf Remarks} }

\def\bel{\begin{equation}\label{} }
\def\om{\omega}
\def\omb{\bar \omega}
\def\qi{\mathfrak{i}}
\def\qj{\mathfrak{j}}
\def\qk{\mathfrak{k}}
\def\uz{\underline{0}}
\def\uone{\underline{1}}
\def\uw{\underline{\omega}}
\def\uwb{\underline{\bar \omega}}

\title{Moonshine, Superconformal Symmetry, and Quantum Error Correction }

\author[1]{Jeffrey A. Harvey,}
\author[2]{Gregory W.~Moore}

\affiliation[1]{Enrico Fermi Institute and Department of Physics\\
$~~ $University of Chicago \\
$~~ $5620 Ellis Ave., Chicago IL 60637}
\affiliation[2]{NHETC and
$~~$Department of Physics and Astronomy, Rutgers University \\
$~~$126 Frelinghuysen Rd., Piscataway NJ 08855, USA }

\emailAdd{j-harvey@uchicago.edu}
\emailAdd{gmoore@physics.rutgers.edu}

\abstract{
Special conformal field theories can have symmetry groups which are interesting sporadic finite
simple groups. Famous examples include the Monster symmetry group of a $c=24$ two-dimensional conformal field theory (CFT) constructed by Frenkel, Lepowsky and Meurman, and the Conway symmetry group of a $c=12$ CFT explored in detail by Duncan and Mack-Crane.
The Mathieu moonshine connection between the K3 elliptic genus and the Mathieu group $M_{24}$ has led to the study of K3 sigma models with large symmetry groups. A particular K3 CFT with a maximal symmetry group preserving $(4,4)$ superconformal symmetry was studied in beautiful
work by Gaberdiel, Taormina, Volpato, and Wendland \cite{gtvw}.  The present paper shows that in both the GTVW and $c=12$ theories the construction of  superconformal generators can be understood via the theory of quantum error correcting codes. The automorphism groups of these codes lift to symmetry groups in the CFT preserving the superconformal generators. In the case of the $N=1$ supercurrent of the GTVW model our
result, combined with a result of T. Johnson-Freyd implies the symmetry group is the maximal subgroup of $M_{24}$ known as the sextet group.
(The sextet group is also known as the  holomorph of the hexacode.)   Building on   \cite{gtvw} the Ramond-Ramond sector of the
GTVW model is related to the Miracle Octad Generator which in turn leads to a role for the Golay code as a group of symmetries of
RR states. Moreover, $(4,1)$ superconformal symmetry suffices to define and decompose the elliptic genus of a K3 sigma model into characters of the $N=4$ superconformal algebra. The symmetry group preserving $(4,1)$ is larger than that preserving $(4,4)$.    
\vskip 0,1in
\today}


\begin{document}
\maketitle

\addtocontents{toc}{\protect\enlargethispage{35mm}}

\section{Introduction}\label{sec:Intro}

Physicists who wonder about the ultimate structure of our universe can be divided into two classes. The first class believes that our universe
is chosen at random from a huge ensemble of possible universes. The second class believes that our universe and the fundamental laws that
govern it are based on some beautiful and exceptional mathematical structure. String theory provides evidence for both points of view.
Anomaly cancellation \cite{Green:1984sg}, the discovery of the heterotic string \cite{Gross:1984dd, Gross:1985fr, Gross:1985rr} and the discovery of Calabi-Yau compatifications \cite{Candelas:1985en} of string theory certainly
involved beautiful and exceptional structures and connections between mathematical objects and the demands of physical consistency.
Subsequent developments involving new constructions of Calabi-Yau spaces, the understanding of D-branes in string theory \cite{Polchinski:1995mt} and the multitudinous possibilities of (alleged) flux compactifications \cite{Douglas:2006es} have led to the idea of a huge landscape of string vacua and no clear principle that would select one
special point in this landscape as describing our universe. The authors of the present paper are philosophically inclined towards the point of view of the second class of physicists and are thus interested in studying special points in the landscape of string compactifications that involve beautiful and
exceptional mathematical structures.  Certainly among these special points the ones associated to moonshine are amongst the most
beautiful. The $c=24$ CFT underlying Monstrous Moonshine can be used to describe a compactification of the bosonic string to two
dimensions. Similarly, the $c=12$ SCFT with Conway symmetry can be used to describe compactification of type II superstring theory
to two dimensions. Moreover, the  holomorphic $c=24$ Monster CFT and an anti-holomorphic version of the $c=12$ Conway theory can be used
to describe a compactification of the heterotic string to two dimensions.  Of course two dimensions is not a good description of the real world. More recently, moonshine
phenomenon that are less well understood but closer to physical reality have been observed in the study of the elliptic genus of $K3$ sigma models \cite{Eguchi:2010ej}. These sigma models provide a compactification of superstring theory to six spacetime dimensions and play a central role in various string dualities in four, five and six spacetime dimensions. Despite ten years of persistent effort by a small but devoted
community of physicists and mathematicians the Mathieu (and Umbral) moonshine phenomena remain mysterious, hinting at an important gap in
our understanding of symmetries of conformal field theories and string models.

Our motivations for the investigations in this paper were multifold. We wanted to understand the origin of moonshine for the sporadic group $M_{24}$ in the elliptic genus of $K3$ and hoped that study of symmetries preserving $(4,1)$ superconformal symmetry might shed some
light on this puzzle. We were also intrigued by our realization that  error correcting codes appear naturally in the analysis of the special K3 sigma model discussed in \cite{gtvw}. We will henceforth refer to this model as the GTVW model in recognition of the analysis in \cite{gtvw} but elements of the model were actually constructed and discussed earlier \cite{Wendlandphd, Wendland:2001ks, Nahm:1999ps}.
As will become clear in later sections, we still do not understand the origin of
$M_{24}$ but we were able to enlarge the possible symmetry transformations that can act on the $K3$ elliptic genus beyond those previously
considered.

Connections between classical error correcting codes and CFT have a long history. Many special lattices can be constructed from classical error correcting codes \cite{splag}
and of course from a lattice one can construct a lattice VOA leading to a connection between special codes and special VOAs, see for example \cite{Dolan:1989kf,Goddard:1989dp}.
There are also code VOAs which utilize classical codes to embed nontrivial CFTs into tensor products of simpler CFTs 
\cite{dgh,junla,lam,miya_bincode}. However as far
as we are aware the role we have found for  {\it quantum error correcting codes} in understanding the structure of superconformal field theories is new.
Our study of codes in the GTVW  model provides a nice example where both classical and quantum error correcting codes play important roles. In particular the classical hexacode
plays a central role in understanding the symmetries
of that model while  the construction of superconformal generators in both the GTVW model  and the $c=12$ Conway Moonshine CFT are linked to
special quantum error correcting codes.
We suspect that there is also such a connection for the superconformal generator in the version of the Monster CFT studied in \cite{Dixon:1988qd} but there are many subtleties in the analysis which have kept us from presenting such an analysis here.

Throughout the text we indicate several directions where further research might be
informative and useful. In addition to these, our work clearly raises the interesting
question of what is the moduli space of $(4,1)$ sigma models smoothly connected to $(4,4)$
K3 sigma models. In particular, one would like to know if there is an analog of the Quantum Mukai
Theorem of Gaberdiel, Hohenegger, and Volpato in this context.  For some literature on $(4,1)$ sigma
models see \cite{Howe:1987qv,Hull:1986hn,Hull:2016khc,Hull:2018jkr}. In \cite{Johnson-Freyd:2019wgb}
$N=1$ superconformal structures on a large class of VOA's were classified. If the main message of
this paper is correct, there should be connections to quantum error-correcting codes in these
examples. It will be interesting to see if that turns out to be the case.

The outline of this paper is as follows. Following this introduction, the second section collects a number of useful facts regarding the field $\IF_4$, the hexacode and the relation of these to the Golay code via the Miracle Octad Generator (MOG). We also briefly discuss quantum error correcting codes and highlight a particular one Qbit quantum error correcting code that will play a role in our later analysis.  In section three we present one of our main results. We recall the GTVW model, and explain that the supercurrents in this model are related to
quantum codes. Section four reviews some aspects of Mathieu moonshine and explains the relevance to this paper.  The Mathieu moonshine observation regarding the sporadic group $M_{24}$ and the elliptic genus of K3 sigma models is based on a decomposition of the elliptic genus into characters of the $N=4$ superconformal algebra \cite{Eguchi:2010ej}.
We point out that this decomposition depends only on the existence of $(4,1)$ superconformal symmetry and that the group of symmetry transformations preserving $(4,1)$ superconformal symmetry is in general larger than the possible groups of $(4,4)$ preserving transformations classified in \cite{Gaberdiel:2011fg}.  Section five uses
the relation of the supercurrents to codes to study the symmetry groups of the GTVW model that preserve various superconformal structures.   We focus on the particular K3 sigma model of \cite{gtvw} and find a role for the Golay code in this model, namely there is an isomorphism between the Golay code and a certain stabilizer group of a left-right symmetric $N=1$ supercharge.
Four Appendices summarize background material,  supersymmetry conventions and some technical details.

For some reviews of moonshine we can recommend \cite{Gannon:2004xi,Duncan:2014vfa}
for mathematically inclined readers and \cite{Kachru:2016nty,Anagiannis:2018jqf} for physicists.

\section*{Acknowledgements}

We thank D.~Allcock, A.~Banerjee, D.~Freed, I.~Frenkel, D.~Harlow,
T.~Johnson-Freyd, C.~Hull, T.~Mainiero, D.~Morrison, S.-H.~Shao,  A.~Taormina and K.~Wendland for helpful conversations and correspondence.
We especially thank M.~ Gaberdiel, T.~Johnson-Freyd, A.~ Taormina, and K.~ Wendland for numerous useful and important remarks on the draft.
 We gratefully acknowledge the hospitality of the Aspen Center for Physics  (under
NSF Grant No. PHY-1066293) where this work was initiated. Part of this work was also done at
 the Stanford Institute for Theoretical Physics, and GM would like to thank the SITP, and
 Shamit Kachru in particular, for hospitality. We have made use of the GAP package for finite group theory in our analysis \cite{GAP} as well as the Magma Computer Algebra system which was
made available through a grant from the Simons Foundation. JH acknowledges support from the NSF \footnote{Any opinions, findings, and conclusions or recommendations expressed in this material are those of the author(s) and do not necessarily reflect the views of the National Science Foundation.} under grant PHY 1520748 and from the Simons Foundation (\#399639).
GM is supported by the DOE under grant
DOE-SC0010008 to Rutgers.

\section{Codes And Error Correction}

Codes, both classical and quantum, play a central role in our analysis. In this section we give the basic definitions of classical and quantum error correcting codes
and provide examples that will play a role in our later analysis.  Useful references include \cite{Gottesman, SPLAG, beth, Mermin, preskill}.

Let $q$ be a prime power. A \emph{classical  q-ary code $\CC$ of length $n$} is a set of vectors in the vector space $\IF_q^n$.
The vectors in the set $\CC$ are called \emph{codewords}: They are words with   $n$ letters drawn from the finite field $\IF_q$,
regarded as an alphabet.
If $\CC\subset \IF_q^n$ is a linear subspace  it is called a \emph{linear code}. The main examples
we encounter in this paper are   linear codes with $q=2$ and $q=4$.    The dimension $k$ of a linear code is the dimension of the linear subspace
spanned by the codewords. Such codes are said to be of type $[n,k]$.  A code of type $[n,k]$
can be specified by a generator matrix $G$ which is a $k \times n$ matrix such that the code ${\cal C}$ is spanned by the rows of $G$ with
coefficients in $\IF_q$. One can always find a code equivalent to ${\cal C}$ such that the generator matrix takes the form
\be \label{genmat}
G=[I_k, A]
\ee
with $A$ a $k \times (n-k)$ matrix and $I_k$ the $k \times k$ identity matrix.
We use the convention that code words $c$ (of length $n$) encode the message $m$ (of length $k$) as
\be
c=mG
\ee
so that with $G$ in canonical form the first $k$ letters of $c$ are simply the elements of the message word $m$.

In the field $\IF_q$ one has the conjugation (Frobenius) automorphism $x \rightarrow \bar x = x^p$ for $q=p^\alpha$. The dual or orthogonal code to a code ${\cal C}$ is
defined to be
\be
{\cal C}^\perp = \{ x \in \IF_q^n: x \cdot \bar u=0, {\rm ~all~}u \in {\cal C} \}
\ee
The dimension of ${\cal C}^\perp$ is $n- {\rm dim} ~{\cal C}$. The code is called self-dual if ${\cal C}= {\cal C}^\perp$.
A useful fact is that if ${\cal C}$ has generator matrix given by eqn. \ref{genmat} then a generator matrix for ${\cal C}^\perp$ is
\be
H = [ - \bar A^{tr}, I_{n-k} ] \, .
\ee
$H$ is also called the parity check matrix of the code ${\cal C}$.

The (Hamming) distance $d(c_1, c_2)$ between any two code words $c_1$, $c_2$ is the number of places at which $c_1$ and $c_2$
differ. The Hamming distance of a code is the minimum distance between any two codewords,
\be
d({\cal C})= \substack{ {\rm min}  \\ c_1, c_2 \in {\cal C}, c_1 \ne c_2}  d(c_1, c_2)
\ee
The Hamming distance is often included in the specification of a code by writing $[n,k,d]$ for a $[n,k]$ code with Hamming distance $d$.

We now give a number of examples of codes that play a role in our later analysis.

\subsection{Codes Related To The Hamming Code}\label{subsec:RelatedHamming}

The codes which govern the $N=1$ superconformal symmetries in a K3 sigma model
we study below are closely related to the renowned Hamming code.

The Hamming code is a binary linear $[7,4,3]$ code over $\IF_2$ with generator matrix.
\be
G=\begin{pmatrix} 1 & 0 & 0 & 0 & 1 & 1 & 0 \\
			     0 & 1 & 0 & 0 & 1 & 0 & 1 \\
			     0 & 0 & 1 & 0 & 0 & 1 & 1 \\
			     0 & 0 & 0 & 1 & 1 & 1 & 1
			     \end{pmatrix} \, .
\ee
The Hamming code can famously detect and correct any single bit error. Adding a parity bit to the code gives the $[8,4]$ Hamming code which
can also detect (but not correct) errors in two bits.

If we simply drop the seventh bit of the Hamming code we obtain instead a $[6,4]$ binary linear code. Relabelling the entries as
$\{x_1+x_2+x_4,x_2,x_1+x_3+x_4,x_3,x_4,x_1 \}$ we obtain a code $\CS_{N=1}$ of type $[6,4]$ whose sixteen codewords can be listed as:
\be \label{eq:sixfourcode}
\begin{split}
[\emptyset], [123456], [1234], [3456], [1256], [12], [34], [56],  \\
[135], [245], [236], [146], [246], [235], [136], [145]  \, .
\end{split}
\ee
Here we are using a notation where we list only the entries that are $1$ for each word.
We have adopted this specific choice in order to facilitate comparison to later expressions. See e.g.
equation \eqref{eq:BasicSpinor}.

A subcode $\CS_{N=4}$ of this truncated Hamming code will also play a role in
describing $N=4$ supercurrents.  $\CS_{N=4}$ is a code of type $[6,3]$ spanned by
\be\label{eq:NewSixThreeCode}
\begin{split}
w_1 &= (0,0,1,1,1,1):=~~[3456] \\
w_2 &= (0,1,0,1,0,1):=~~[246] \\
w_3 & =(0,0,0,0,1,1):=~~[56] \\
\end{split}
\ee

\subsection{The Hexacode}\label{subsec:hexacode}

The hexacode utilizes the field $\IF_4$.
Since some properties of this field might not be familiar to physicists we briefly review them here.
We think of it concretely as the set
\be
\IF_4 = \{ \uz, \uone,\uw,\uwb \} \, .
\ee
To define the Abelian group law for addition of vectors we take $\uz$ to be the additive identity and
\be
\begin{split}
\uone+ \uw & = \uwb \\
\uone + \uwb & = \uw \\
\uw + \uwb & = \uone \\
x + x & = \uz \, .
\end{split}
\ee
We write $\IF_4^{+}$ for $\IF_4$ considered as an Abelian group with the $+$ law.
As an Abelian group it is isomorphic to $\IZ_2 \oplus \IZ_2$.
Note that $\IF_4$ is a field of characteristic two, so $2x=\uz$ for all $x$,
and there is no distinction between $x$ and $-x$.

The multiplicative Abelian group law for nonzero vectors $\IF_4^*$ is defined
by taking  $\uone$ to be the multiplicative identity and
\be
\begin{split}
\uw \uw & = \uwb \\
\uw \uwb & = \uone \\
\uwb \uwb & = \uw \\
\end{split}
\ee

Recall that  $\mathbb{F}_4$ has a $\IZ_2$ (Frobenius) automorphism $x\to x^2$. This automorphism preserves
the additive and multiplicative identities $\uz$ and $\uone$ and takes
takes
\be
\uw \leftrightarrow \uwb   \, .
\ee

In our analysis of the GTVW K3 sigma model an important role is played by a group homomorphism
from the quaternion subgroup of $SU(2)$ to $\IF_4^+$. To define this we first
consider the quaternion group
$Q \subset SU(2)$ generated by $ \I \sigma^1, \I \sigma^2, \I \sigma^3$ where the $\sigma^i$ are the standard Pauli matrices. This is an
$8$ element group. Explicitly:
\be\label{eqn:quatgroup}
Q = \{ \pm 1 , \pm \I \sigma^1, \pm \I \sigma^2, \pm \I \sigma^3 \}
\ee
with group composition given by matrix multiplication. There is a homomorphism to $\IF_4^{+}$ with kernel given by the subgroup $\{ \pm 1\}$:
\be\label{eq:QtoF4ab}
1 \rightarrow \{ \pm 1\} \rightarrow Q ~ {\buildrel \pi \over \rightarrow } ~ \IF_4^{+} \rightarrow 1
\ee
The sequence does not split.
We will make a specific choice of section $h$ of $\pi$ in our computations:
\be\label{eq:Def-of-h}
\begin{split}
h(\uz) & = \begin{pmatrix} 1 & 0 \\ 0 & 1 \\ \end{pmatrix} \\
h(\uone) & = \begin{pmatrix} 0 & 1 \\ -1 & 0 \\ \end{pmatrix} = i \sigma^2\\
h(\uw) & = \begin{pmatrix} 0 & \I  \\ \I  & 0 \\ \end{pmatrix}   = i \sigma^1 \\
h(\uwb) & = \begin{pmatrix} -\I  & 0 \\ 0  & \I  \\ \end{pmatrix} =-i \sigma^3 \, . \\
\end{split}
\ee
There is also a relation of the multiplicative structure of $\IF_4$ with the quaternion group:
\be\label{eq:OmegaConj}
\begin{split}
h(\uw x) & = h(x \uw) = \Omega^{-1} h(x) \Omega \\
h(\uwb x) & =h(x \uwb)= \Omega^{-2} h(x) \Omega^{2} = \Omega h(x) \Omega^{-1} \, . \\
\end{split}
\ee
Note the second equation immediately follows from the first since
\be
h(\uwb x) = h(\uw \uw x) = \Omega^{-1} h(\uw x) \Omega = \Omega^{-2} h(x) \Omega^2
\ee
In the above
\be
\Omega = \begin{pmatrix} \half(1-\I) & \half (1+\I) \\   - \half(1-\I) & \half (1+\I) \\ \end{pmatrix}
=\frac{1}{\sqrt{2}} \begin{pmatrix} 1 &  1 \\ -1 & 1 \\ \end{pmatrix}
\begin{pmatrix} e^{-\I \pi/4} & 0 \\ 0 & e^{\I \pi/4} \\ \end{pmatrix}
\ee

We collect here a few useful properties of $\Omega$
\begin{enumerate}
\item $\Omega^3 = -1$ so $\Omega^{-1} = - \Omega^2$.
\item $\Omega$ is an $SU(2)$ matrix. Its action on $\IR^3$ is an order three
rotation around the axis through $(1,1,1)$ so it permutes $x\to y\to z\to x$.
This mirrors the action of multiplication by $\omega$ which permutes the
$3$ elements of $\IF_4^*$:
\be
\xymatrix{  & \uone \ar[ld]  &  \\
\uw \ar[rr] & & \uwb  \ar[lu] }
\ee

\end{enumerate}

Returning to \eqref{eq:Def-of-h}, we can define a $\IZ_2$-valued cocycle
\be\label{eq:h-cocycle-def}
h(x) h(y) = \epsilon(x,y) h(x+y)
\ee
for $x,y\in \IF_4$. This cocycle has some nice properties:
\begin{enumerate}
\item It is $1$ if $x$ or $y$ is the additive identity,
\be
\epsilon(0,x) = \epsilon(x,0) = 1 \, .
\ee
\item The diagonal values are
\be
\epsilon(x,x) = \begin{cases}
+1 & x=\uz \\
-1 & x\not= \uz \, .
\\
\end{cases}
\ee
\item $\epsilon$ is ``permutation invariant''
\be\label{eq:CocycPermInv}
\epsilon(\uw x , \uw y) = \epsilon(x,y) \, .
\ee
\item  $\epsilon$ is bimultiplicative:
\be
\epsilon(x_1+x_2,y) = \epsilon(x_1,y) \epsilon(x_2,y)\, ,
\ee
\be
\epsilon(x,y_1+y_2) = \epsilon(x,y_1) \epsilon(x,y_2) \, .
\ee
\item For $x,y\not= \uz$ we have, from \eqref{eq:Def-of-h},
\be
\epsilon_{x,y} =
\begin{pmatrix}
-1 & -1 & 1 \\
1 & -1 & -1 \\
-1 & 1 & -1 \\
\end{pmatrix} \, .
\ee
\end{enumerate}

To prove permutation invariance \eqref{eq:CocycPermInv}
 we simply conjugate \eqref{eq:h-cocycle-def}
by $\Omega$ and use \eqref{eq:OmegaConj}. To prove the
bimultiplicative property we first note that it is obvious
if any argument is zero, so we can assume all arguments
are nonzero. We are not aware of any better proof than an
explicit check of the   9 cases to estabish
$\epsilon(x_1+x_2,y) = \epsilon(x_1,y) \epsilon(x_2,y) $.
Then one uses the symmetry properties of $\epsilon$.

We now turn to a discussion of the hexacode $\CH_6$, a three-dimensional
subspace of $\IF_4^6$ consisting of codewords $(x_1,\dots, x_6)$
such that, if
\be
\Phi_{x_1,x_2,x_3}(y) := x_1 y^2 + x_2 y + x_3
\ee
then
\be\label{eq:quadfundef}
\begin{split}
x_4 & = \Phi_{x_1,x_2,x_3}(\uone) \\
x_5 & = \Phi_{x_1,x_2,x_3}(\uw) \\
x_6 & = \Phi_{x_1,x_2,x_3}(\uwb) \\
\end{split}
\ee
A full list of the 64 hexacode words (which can occasionally be quite
helpful when doing computations) is shown in
Appendix \ref{app:HexacodeAutGroup} where we also analyze the automorphism group of the hexacode.

Note that $\Phi_{x_1,x_2,x_3}$ depends linearly on $x_1,x_2,x_3$ so this
makes manifest that we have a three-dimensional subspace. In fact, taking
$(x_1,x_2,x_3) = (\uone,\uz,\uz), (\uz,\uone,\uz),(\uz,\uz,\uone)$  we automatically
get a basis
\be\label{eq:hexbasis}
\begin{split}
b_1 & = (\uone,\uz,\uz,\uone,\uwb,\uw ) \\
b_2 & = (\uz,\uone,\uz,\uone,\uw, \uwb ) \\
b_3 & = (\uz,\uz,\uone,\uone,\uone,\uone) \\
\end{split}
\ee
corresponding to the generator matrix
\be
\begin{pmatrix}
\uone & \uz & \uz & \uone & \uwb & \uw \\
\uz & \uone & \uz & \uone & \uw & \uwb \\
\uz & \uz & \uone & \uone & \uone & \uone \\
\end{pmatrix} = \begin{pmatrix} 1 , & A \\ \end{pmatrix}
\ee

We can make an inner product space with a nondegenerate $\IF_4$-valued inner product:
\be
\langle x, y \rangle := \sum_{i=1}^6  \bar x_i y_i
\ee

There is nice relationship between the hexacode and the $[6,4]$ binary linear code defined earlier. To see this we first define
a homomorphism of Abelian groups
\be\label{eq:pi-projection}
\pi: \IF_4^{+} \to \IF_2
\ee
such that
\be\label{eq:pi-projection-def}
\begin{split}
\pi(\uz) =  0 \qquad & \qquad \pi(\uwb) = 0 \\
\pi(\uone) =  1 \qquad & \qquad \pi(\uw) = 1 \, . \\
\end{split}
\ee
The conceptual reason behind this choice of $\pi$ is related to the
quantum code interpretation discussed below: Operators $h(x)$ that do not
induce a bit-flip map to $0$ and those that induce a bit-flip map to $1$.
We can extend this in an obvious way to
\be
\pi: \left( \IF_4^{+} \right)^6 \to \IF_2^6 \, .
\ee
Applying $\pi$ to the 64 code words of the hexacode clearly produces the $16$ words of a binary linear $[6,4]$ code and a short computation
shows that the resulting $[6,4]$ code is isomorphic to the one defined in equation  \eqref{eq:sixfourcode}.

The dual, or orthogonal, code $\CH_6^*$ is defined by the set of
$x$ so that $\langle y, x \rangle=0$ for all $y\in \CH_6$. It is not
hard to check that the hexacode is self-dual: It is maximal isotropic.

It is useful to note another maximal isotropic subspace $\CH_6^\perp$
so that we have a decomposition into maximal isotropic subspaces:
\be\label{eq:MaxIsotropicDecomp}
\IF_4^6  \cong \CH_6 \oplus \CH_6^{\rm comp}
\ee
We can do this as follows. Let $e_i$ be the vector with
a $\uone$ in the $i^{th}$ coordinate and $\uz$ elsewhere.
Clearly we have
\be
\langle e_i, b_j \rangle = \delta_{ij}\qquad \qquad 1\leq i,j \leq 3
\ee
But the $e_i$ do not span an isotropic subspace. Nevertheless, since the hexacode is
isotropic we can try to modify the span of $e_i$, $i=1,2,3$ to the span of
\be
u_i = e_i + A_{ij} b_j
\ee
 and then the desired
\be
\langle u_i, u_j \rangle = \uz
\ee
implies that
\be
\delta_{ij} + \bar A_{ij} + A_{ji} = \uz
\ee
There are many solutions to this, but a simple one is to take
$A_{ij} = \uw \delta_{ij}$. So if
\be
\begin{split}
u_1 & = (\uwb, \uz,\uz,\uw,\uone,\uwb ) \\
u_2 & = (\uz,\uwb,\uz,\uw,\uwb, \uone ) \\
u_3 & = (\uz,\uz,\uwb,\uw,\uw,\uw) \\
\end{split}
\ee
then
\be
\langle b_i, u_j \rangle = \delta_{ij} \qquad   \langle u_i, u_j \rangle = 0
\ee
so the span of $u_i$ will serve as a maximal isotropic complementary space $\CH_6^{\rm comp}$.

\subsection{The Hexacode And The Golay Code}

In this section we rephrase slightly the description of the
``Miracle Octad Generator'' (MOG) \cite{curtis}  which can be found in
\cite{SPLAG}. This is a particularly efficient way of
thinking about the Golay code. In \cite{gtvw} the MOG was connected to the space of R-R ground states of a particular K3 sigma model
and our later analysis will elaborate on, and clarify,  this connection.

The MOG is based on a map from digits in $\IF_4$ to $\IF_2^4$.
There are four ``interpretations'' of the digits and these
can be summarized by a pair of maps
\be
g^\pm: \IF_4 \times \IF_2 \rightarrow \IF_2^4
\ee

The map $g^+$ corresponds to what SPLAG \cite{SPLAG} refers to as  ``even interpretation of hexacode digits''
and is defined by:
\be\label{eq:fg-LR-1-app}
g^+(\underline{0},0) = \begin{pmatrix} 0 \\ 0 \\ 0 \\ 0 \\  \end{pmatrix} \quad
g^+(\underline{0},1) = \begin{pmatrix} 1 \\ 1 \\ 1 \\ 1 \\ \end{pmatrix} \quad
g^+(\underline{1},0) = \begin{pmatrix} 0 \\ 0 \\ 1 \\ 1 \\ \end{pmatrix} \quad
g^+(\underline{1},1) =  \begin{pmatrix} 1 \\ 1 \\ 0 \\ 0 \\ \end{pmatrix}\quad
\ee
\be\label{eq:fg-LR-2-app}
g^+(\underline{\omega},0) = \begin{pmatrix} 0 \\ 1 \\ 0 \\ 1 \\ \end{pmatrix}  \quad
g^+(\underline{\omega},1) = \begin{pmatrix} 1 \\ 0 \\ 1 \\ 0 \\  \end{pmatrix} \quad
g^+(\underline{\bar\omega},0) = \begin{pmatrix} 0 \\ 1 \\ 1 \\ 0 \\  \end{pmatrix} \quad
g^+(\underline{\bar\omega},1) = \begin{pmatrix} 1 \\ 0 \\ 0 \\ 1 \\ \end{pmatrix} \quad
\ee
Note that $g^+$ is in fact a group homomorphism $\IF_4^+ \times \IF_2 \to \IF_2^4$.

Now we define a group homomorphism
\be
\mathfrak{f}^+: (\IF_4^+ \times \IF_2)^6 \to \IF_2^{24}
\ee
It is very useful to represent the vectors in $\IF_2^{24}$ as a $4\times 6 $
array of elements of $\IF_2$. Each column is the result of $g^+$ applied to
a ``decorated digit'' in $\IF_4 \times \IF_2$. That is,
\bel
\mathfrak{f}^+(x_1,\epsilon_1, \dots, x_6, \epsilon_6 ) = ( g^+(x_1,\epsilon_1), \dots, g^+(x_6, \epsilon_6) ) \qquad
\qquad x_\alpha \in \IF_4\quad \epsilon_\alpha\in \IF_2
\ee
See Table \ref{mogtab} for an illustration of this construction. We will refer to
elements
\be\label{eq:DecoratedWord}
(x_1,\epsilon_1, \dots, x_6, \epsilon_6 )  \in (\IF_4 \times \IF_2)^6
\ee
as \emph{decorated words}, viewing $w = (x_1, \dots, x_6) \in \IF_4^6$ as the
word and $(\epsilon_1, \dots, \epsilon_6)$ as the decorations. We also
denote decorated words as $(x,\epsilon)$.

\begin{table}
\begin{center}
\begin{tabular}{|c|c|c|c|c|c|}
\hline
0 & 1 & 0 & 1 & 1 & 1  \\
0 & 1 & 0 & 1 & 0 & 0 \\
0 & 0 & 0 & 0 & 1 & 0 \\
0 & 0 & 0 & 0 & 0 & 1 \\
\hline
$\uz$ & $\uone$ & $\uz$ & $\uone$ & $\uw$ & $\uwb $ \\
\hline
\end{tabular}
\end{center} \caption{Illustration of the MOG construction of a Golay code word from an even interpretation of the hexacode word $\uz \uone \uz \uone \uw \uwb$ which is displayed along the bottom of the table.} \label{mogtab}
\end{table}

The first nontrivial fact is that the  image of $\mathfrak{f}^+$ on the decorated words
$(x,\epsilon)$ with
\be\label{eq:GolayEven1}
x\in \CH_6 \qquad\qquad
\sum_{\alpha=1}^6 \epsilon_\alpha = 0 ~ \mod ~ 2
\ee
is an index two subspace of the Golay code $\CG \subset \IF_2^{24}$. We will call it the
 \emph{even Golay code} and denote it by  $\CG^+$. This is a set of $2^{11}$ vectors and forms a
subgroup of the Golay code because $g^+$ and $\mathfrak{f}^+$ are homomorphisms.

The complement of the set of even Golay code words inside the set of all Golay code words is the set of
odd Golay code words. A good way to parametrize the odd Golay code words is to introduce the
column vector
\be
p = \begin{pmatrix} 1\\ 0 \\ 0 \\ 0 \\  \end{pmatrix}
\ee
and we then define
\be
g^-(x,\epsilon):= g^+(x,\epsilon) + p
\ee
so
\be\label{eq:fg-LR-1-app-minus}
g^-(\underline{0},0) = \begin{pmatrix} 1 \\ 0 \\ 0 \\ 0 \\  \end{pmatrix} \quad
g^-(\underline{0},1) = \begin{pmatrix} 0 \\ 1 \\ 1 \\ 1 \\ \end{pmatrix} \quad
g^-(\underline{1},0) = \begin{pmatrix} 1 \\ 0 \\ 1 \\ 1 \\ \end{pmatrix} \quad
g^-(\underline{1},1) =  \begin{pmatrix} 0 \\ 1 \\ 0 \\ 0 \\ \end{pmatrix}\quad
\ee
\be\label{eq:fg-LR-2-app-minus}
g^-(\underline{\omega},0) = \begin{pmatrix} 1 \\ 1 \\ 0 \\ 1 \\ \end{pmatrix}  \quad
g^-(\underline{\omega},1) = \begin{pmatrix} 0 \\ 0 \\ 1 \\ 0 \\  \end{pmatrix} \quad
g^-(\underline{\bar\omega},0) = \begin{pmatrix} 1 \\ 1 \\ 1 \\ 0 \\  \end{pmatrix} \quad
g^-(\underline{\bar\omega},1) = \begin{pmatrix} 0 \\ 0 \\ 0 \\ 1 \\ \end{pmatrix} \quad
\ee

Now, the odd Golay code words are
\be\label{eq:odd-golay-words}
w= \left(  g^-(x_1,\epsilon_1), \dots,  g^-(x_6, \epsilon_6) \right)
\ee
with
\be
\sum_{\alpha =1}^6 \epsilon_\alpha = 1~ \mod~ 2 ~ .
\ee

The MOG construction of the Golay code gives a nice connection between the automorphism group
of the even Golay code $\CG^+$ and the holomorph of the hexacode. (See Appendix \ref{app:HexacodeAutGroup} for the
definition of the holomorph of a group.)
The hexacode itself is $2^6$ as an Abelian group and its automorphism group
is described in Appendix \ref{app:HexacodeAutGroup}. The holomorph has the structure $2^6:3.S_6$.
The hexacode acts by
\be
w\cdot \ff^+(x,\epsilon) = \ff^+(x+w,\epsilon)
\ee
for $w\in \CH_6$ and
\be
\varphi\cdot \ff^+(x,\epsilon) = \ff^+(\varphi^{-1}(x),(p(\varphi))^{-1}\cdot \epsilon)
\ee
where $\varphi \in \Aut(\CH_6)$ and $p$ is defined in equation \eqref{eq:TrackPerm}.
The automorphism group of $\CG^+$
within $S_{24}$ can be shown to be exactly the maximal subgroup of
$M_{24}$ known as the sextet group.
\footnote{We thank D. Allcock for explaining this to us.}
The sextet group is nicely described in \cite{SPLAG} (see also the
very informative Wikipedia article on $M_{24}$). In fact the sextet
group is exactly the same as the holomorph of the hexacode.

\subsection{Quantum Codes}

Quantum error correcting (QEC) codes are designed to detect and correct errors in the transmission and processing of quantum information.
The set of error operations $\CE$ is a subset of the space of completely positive, non-trace increasing maps on the space of density matrices.
We define a QEC as follows:

\medskip
\noindent \textbf{Definition}
Let H be a Hilbert space.  A Hilbert subspace  $C \subset H$ is an \emph{error correcting quantum code
with respect to a set $\CE$ of error operations} if, for every $E_i \in \CE$,  we have
\be
P E_i^\dagger E_j P = \alpha_{ij} P
\ee
where $P$ is the projector onto $C$ and  $\alpha_{ij}$ is a nonzero Hermitian matrix.
\medskip

When the above condition is satisfied it is possible to
construct a quantum operation ${\cal R}$ that detects and then corrects the quantum errors $E_i$ while preserving the quantum information
carried by states in the quantum code $C$. In some cases we will consider codes that can only detect but not correct quantum errors.

 The QECs we consider are constructed as subspaces of the
$n$-Qbit Hilbert space ${\cal H}_n=(\IC^2)^{\otimes n}$ and the set of error operations will consist of tensor products of elements $X_i, Y_i, Z_i$
which act on the $i^{th}$ component of ${\cal H}_n$ as the Pauli matrices
\be
X= \begin{pmatrix} 0 & 1 \\ 1 & 0 \end{pmatrix}, ~~~~Y= \begin{pmatrix} 0 & -\I \\ \I & 0 \end{pmatrix},~~~~ Z= \begin{pmatrix} 1 & 0 \\ 0 & -1 \end{pmatrix}
\ee
We refer to these tensor products $E_a$ as Pauli operators and we define the weight of a Pauli operator to be the number of entries
in the tensor product that have a non-trivial Pauli operator ($X,Y,Z$) rather than the identity operator ($I$). Thus
$Z \otimes X \otimes I \otimes I$ has weight two.

If the code subspace of dimension $2^k$ is embedded in ${\cal H}_n$ with $k<n$ we will refer to the code as a quantum code
of type $[[n,k]]$,
where the double square brackets are used to distinguish quantum codes from classical codes as in \cite{preskill}.  The code distance
of a quantum code is the minimum weight of a Pauli operator  in $\CE$ that is not the identity operator on $C$.
If a quantum code is of type $[[n,k]]$ and has code distance $d$ we say it is of type $[[n,k,d]]$.

Stabilizer QEC codes are an important class of quantum codes that have much in common with classical linear codes. In the stabilizer formalism
one considers the Pauli group $G_n$ acting on $n$ Qbits. The Pauli group acting on a single Qbit has a matrix representation consisting of the
Pauli matrices along with multiplicative factors of $\pm1 $ and $\pm \I$. Explicitly
\be
G_1= \{\pm I, \pm \I I, \pm X, \pm \I X, \pm Y, \pm\I Y, \pm Z, \pm \I Z \}
\ee
with $I$ the two by two identity matrix. (The quaternion group $Q$ is the intersection $Q = G_1 \cap SU(2)$:
\be\label{eq:quatgroup}
Q = \{\pm I,    \pm \I X,  \pm \I Y,   \pm \I  Z \}
\ee
and will be of some use later.)

 Elements of the general Pauli group $G_n$ are n-fold tensor products of elements of $G_1$. Let $S$ be
a subgroup of $G_n$ and $V^S$ the subspace of ${\cal H}_n$ stabilized by $S$. (We have $V^S=\{ 0 \}$ if $S$ contains the group element $-1$,
so we avoid this case.) Many important quantum error correcting codes can be constructed
as subspaces of the form $V^S$ for particular choices of $S$.

An important example is the smallest QEC capable of detecting and correcting an arbitrary single one Qbit error, that is
a code with ${\cal E}= \{X,Y,Z \}$.
 This code is of type $[[5,1,3]]$ and can be viewed as a stabilizer code
with stabilizer group $S$  generated by the four elements
\begin{align}
M_1 &= XZZXI \\
M_2 &= IXZZX \nonumber \\
M_3 &= XIXZZ  \nonumber \\
M_4 &= ZXIXZ \nonumber
\end{align}
Here as is usual in the QEC literature
$XZZXI$ is shorthand for $X \otimes Z \otimes Z \otimes X \otimes I$ acting on the 5-Qbit space $(\mathbb{C}^2)^{\otimes 5}$.
Note that $M_{2,3,4}$ are obtained from $M_1$ by cyclic permutation.  An orthonormal basis for the code subspace can be constructed
by starting with $|00000 \rangle$ or $|11111 \rangle$ and summing over all elements of the stabilizer group $S$. This leads to the basis
vectors (see e.g. section 10.5.6. of \cite{nielsen})
\begin{align}\label{eq:StabCodeword}
| 0_L \rangle &= \frac{1}{4} \big( |00000 \rangle + | 10010 \rangle + |01001 \rangle + |10100 \rangle \\
		  &~~~+|01010 \rangle - |11011 \rangle - |00110 \rangle - |11000 \rangle    \nonumber \\
		  &~~~ -|11101 \rangle - |00011 \rangle -|11110 \rangle -|01111 \rangle   \nonumber \\
		  & ~~~ -|10001 \rangle - |01100 \rangle - |10111 \rangle + |00101 \rangle \big)  \nonumber
		  \end{align}
and $|1_L \rangle = X^{\otimes 5} |0_L \rangle$.
There is also a $[[6,0,4]]$ code which is one-dimensional and is constructed from the basis of the $[[5,1,3]]$ code and an ancillary Qbit as
\be\label{eqn:psisixfour}
\Psi_{[[6,0,4]]}=  \frac{1}{\sqrt{2}} \left( |0 \rangle \otimes | \bar 0 \rangle + |1 \rangle \otimes |\bar 1 \rangle \right)
\ee

\bigskip
\noindent
\textbf{Remark}
J. Preskill \cite{preskill}  notes that this code state is maximally entangled in that the density matrix obtained by tracing over any $3$ Qbits is totally random,
$\rho^{(3)} = I/8$.  Strictly speaking this is an error detecting rather than error correcting code
in that it can detect, but not correct, any single Qbit error. In \cite{Mainiero:2019enr} T.~Mainiero defines
several notions of ``entanglement homology,'' a homological
generalization of entanglement entropy which measures correlations between many body operators in a given state. The Poincar\'e polynomial
for $|  0_L \rangle$, which is \emph{a priori} a five-term polynomial in $q$ turns out to be simply $51 q^2$ for
Mainiero's ``GNS cohomology,''
while the Poincar\'e polynomials for other related homologies are similarly quite simple. In other words: The state, while maximally
entangled has unusual correlation properties.

\section{Supercurrents, States, And Codes} \label{sec:QEC}

In this section we show that the spectra and supercurrents of several SCFTs that exhibit Moonshine phenomena can be rephrased in terms
of both classical and quantum error correcting codes.

\subsection{The GTVW Model}\label{subsec:gtvw}

The GTVW model is a $(4,4)$ sigma model with a very specific $T^4/\IZ_2$ target \cite{gtvw}. It is a distinguished
K3 because it has the maximal group $2^8{\,:\,}M_{20}$ allowed by the results of \cite{Gaberdiel:2011fg} (GHV).
(In section \ref{subsec:44Preserving} we will describe the structure of the group in excruciating detail.)  The
target $T^4$ is the $Spin(8)$ maximal torus. Reference \cite{gtvw} shows that there
are several incarnations of this model and the one we find
most convenient is an extension of a product of six $SU(2)$ level 1  WZW models.
(Equivalently, the product of six Gaussian models at the T-duality invariant point
$R=1$.)

Recall that the $SU(2)$ level $1$ WZW model has affine $SU(2)_L \times SU(2)_R$ symmetry
with both factors at level $1$. The unitary highest weight representations of affine level
$1$ $SU(2)$ will be labeled by $\CV_0$ and $\CV_1$ according to whether the states of lowest
$L_0$ value are in the singlet or doublet of the global $SU(2)$ symmetry. The product of
six such models therefore has a Hilbert space of states on the circle given by
\be\label{eq:SU2-HS}
\CH =\left(  \CV_0 \otimes \tilde \CV_0 \oplus \CV_1 \otimes \tilde \CV_1\right)^{\otimes 6}
\ee
where a tilde denotes right-movers. In the GTVW model
we take instead the Hilbert space to be
\bel\label{eq:xGTVW-HS}
\CH^{\rm GTVW} = \oplus_{(s,\tilde s) \in \CS }  \CV_s \otimes \tilde \CV_{\tilde s}
\ee
where $s,\tilde s\in \IF_2^6$ and $\CS\subset \IF_2^6 \times \IF_2^6$ is the
subgroup defined by demanding that the components of $s, \tilde s$ obey $s_\alpha = \tilde s_{\alpha} + x $ for all $1\leq \alpha \leq 6$
where $x\in \IF_2$ is independent of $\alpha$ and can be either $0$ or $1$.  One easily
checks that the space of fields is closed under operator product expansion since
the fusion rules for $\CV_0, \CV_1$ are simply addition in $\IF_2$.
However, the fields are only mutually local up to sign: The GTVW model is not
a standard CFT but an extension with a super-chiral algebra (a super-vertex-operator algebra)
in which the modules all have mutual locality at worst $\pm 1$.
As we will see, the theory of codes pervades this model, and it is worth noting that
the spectrum itself can be described by a code of type $[12,7]$, although we will not
make use of that fact.

It is useful to note that the space of states \eqref{eq:xGTVW-HS} is $(\IF_2 \oplus \IF_2)$-graded.
One grading is provided by $x$  and the other is $y= \sum_\alpha  s_{\alpha}$.
In terms of the original supersymmetric K3 sigma model we can
identify $\CH^{\rm GTVW}$ with the sum of NSNS and RR sectors of the sigma model.
The quantum number $y=0,1$ distinguishes the NS-NS sector from the R-R sector, respectively.
Moreover reference \cite{gtvw} shows that $e^{\I \pi x}$ can be identified with $(-1)^{F_L + F_R}$ where $F_{L,R}$ are
left- and right-moving fermion numbers. Thus, the GSO-projected version of this model, that is,
the GSO projected K3 sigma model is equivalent to the product of six $k=1$ $SU(2)$ WZW models.

\textbf{Remark}: $\IZ_2$-extensions of vertex operator algebras go back to the
very beginning of superstring theory \cite{Ramond:1971gb,NeveuSchwarz}, and play a crucial role
in formulating tachyon-free models \cite{Gliozzi:1976qd}. They have also played
a role in previous investigations into Moonshine  \cite{Dixon:1988qd}.
A modern interpretation of these ``$\IZ_2$ extensions'' of a conformal
field theory is that one is defining a theory
that depends on both conformal structure and spin structure. The procedure is to identity a
nonanomalous $\IZ_2$ global symmetry of the bosonic CFT,  couple
to an invertible topological field theory sensitive to spin structure known as the Arf
theory, and gauge the diagonal global $\IZ_2$ symmetry of the product theory.  The resulting
theory has a super-VOA as its ``chiral algebra.'' The GSO projection of the theory reproduces
the original bosonic CFT.
\footnote{This procedure, which in fact goes back to \cite{AlvarezGaume:1986mi,AlvarezGaume:1987vm}
has also been used to formulate theories of self-dual fields \cite{Belov:2006jd,Witten:1996hc}.
It has recently been explored and extended further in
\cite{Gaiotto:2015zta,Kapustin:2017jrc,Karch:2019lnn,Lin:2019hks}.
See the 2019 TASI lectures by Y. Tachikawa for a pedagogical introduction.
We thank Shu-Heng Shao for guiding us through this recent literature. }
In our case, the original bosonic CFT is the product of six $SU(2)$
level $k=1$ WZW models, and the GTVW model is the spin extension based on the diagonal element $-1^6 \in SU(2)_L^6$
in the $(SU(2)_L \times SU(2)_R)^6$ WZW model.

\subsection{$(4,4)$ Superconformal Algebra In The GTVW Model}

The K3 sigma model has $(4,4)$ supersymmetry. The left-moving
N=4 supercurrents can therefore be expressed in the WZW language.
The currents must be (anti-) holomorphic primary fields of conformal
dimension $3/2$. To give an explicit construction we use the Frenkel-Kac-Segal construction of level one affine $SU(2)^6$ in terms of
six free bosonic fields $X^I$, $I=1,2, \cdots 6$. We then note that for $\epsilon, \tilde\epsilon \in \{ \pm 1 \}^6$ we define (anti-) holomorphic
vertex operators of conformal dimension $6 \times \frac{1}{4} = \frac{3}{2}$.
\be\label{eq:Vepsilon}
V_{\epsilon} = e^{\frac{\I}{\sqrt{2}} \epsilon \cdot X }(z) \qquad \qquad \tilde V_{\tilde \epsilon}
= e^{\frac{\I}{\sqrt{2}} \tilde \epsilon \cdot \tilde X }(\bar z)
\ee
(For simplicity we will drop cocycle factors as they will play no crucial direct role in our analysis below.)
Any linear combination of these operators gives an (anti-) holomorphic primary field so the
space of holomorphic primary fields  of dimension $3/2$ is a linear space of complex dimension $2^6$
 and the $V_{\epsilon}$ provide a natural basis.  We will find it useful to identify this vector space
 with the Hilbert space of six  Qbits. The basis $V_{\epsilon}$ corresponds to the natural spin basis
\be
\vert \epsilon_1 \rangle \otimes \cdots \otimes \vert \epsilon_6 \rangle
\ee
where, for a single tensor factor, $\{ \vert \epsilon=+ \rangle, \vert \epsilon = - \rangle \}$ is an ordered basis for a Qbit in which a
basis of anti-Hermitian generators of $\fs\fu(2)$ has matrix representation $J^i = -\frac{\I}{2} \sigma^i$, $i=1,2,3$.
If $s\in (\IC^2)^{\otimes 6}$ we can write $s=\sum_{\epsilon \in \IZ_2^6} c(\epsilon) | \epsilon \rangle $ and then we define
\be
V_s= \sum_{\epsilon \in \IZ_2^6} c(\epsilon) V_\epsilon
\ee
to be the corresponding primary field.

The results of \cite{gtvw} are easily used to show that the N=4 supercurrents of the K3 sigma model can be written as very
special states in the six Qbit system. Using the conventions of Appendix \ref{app:SuperAlgebras} we can express these currents
as:
\footnote{To compare to \cite{gtvw} we identify, $Q^1 = G^+$, $\bar Q^1 = G^-$, $\bar Q^2 = G^{'+} $, $Q^2 = G^{'-}$.}
\be
\begin{split}
Q^1 = \left( \frac{\I-1}{2} \right) V_{\Psi^1} \qquad & \qquad   \bar Q^1  = \left( \frac{\I-1}{2} \right) V_{\bar \Psi^1}      \\
Q^2  = \left( \frac{\I-1}{2} \right) V_{\Psi^2} \qquad & \qquad  \bar Q^2   = \left( \frac{\I-1}{2} \right) V_{\bar \Psi^2}  \\
\end{split}
\ee
where
\be
\begin{split}
\Psi^1 & =
  [\emptyset] + [3456] - \I [246] - \I [235] + \I [56] + \I [34] + [245] + [236]  \\
\Psi^2 & = - [1] - [13456] + \I [1246] + \I [1235] - \I [156] - \I [134] - [1245] - [1236]  \\
\bar \Psi^1& =  \I [123456] + [1234] + [1256] - \I [136] - \I [145] +[135] + [146] + \I [12]   \\
\bar \Psi^2 & =  \I [ 23456] + [ 234] + [ 256] - \I [ 36] - \I [ 45] +[ 35] + [ 46] + \I [ 2]  \\
\end{split}
\ee
The notation here is the following: The integers denote the position of a down-spin in the tensor product
of six up/down spins, all other spins being up. Thus, for example,
\be
\begin{split}
[\emptyset] & = \vert + + + + + + \rangle \\
[3456] & = \vert + + - - - - \rangle \\
[145] & = \vert - + + - - + \rangle \\
\end{split}
\ee
and so forth.

For reasons that will be explained later, we are particularly interested in automorphisms of
the K3 sigma model preserving one holomorphic N=1 supercurrent.
If we impose a unitarity constraint then, up to a general $SU(2)_R$ symmetry
transformation the general $N=1$ current is proportional to $Q^1 + \bar Q^1$,
which in turn is proportional to $V_{\Psi}$ with
\be\label{eq:BasicSpinor}
\begin{split}
\Psi & := [\emptyset] +  \I [123456] + ( [1234] +  [3456]+ [1256] ) +    \I ( [12] + [56] +  [34])  \\
& + ([135] + [245] + [236] + [146]) - \I ([246] + [235] +  [136] +  [145] ) \\
\end{split}
\ee
We will show in the next section that this seemingly complicated state
is in fact governed in a simple way by a quantum error detecting code on six Qbits.

\subsection{The Hexacode Representation Of The $N=1$ Supercurrent}\label{subsec:HexaRepSuperCurrent}

We are going to use the hexacode  discussed in section \ref{subsec:hexacode} to construct a rank one projector $P$ on
the Hilbert space of six Qbits. Then our state $\Psi$ will be a vector in the
image of $P$.

The section $h:\IF_4 \to Q \subset SU(2)$ defined in equation
\eqref{eq:Def-of-h} can be extended to vectors in $\IF_4^6$.
If $w= (x_1, \dots, x_6) \in \IF_4^6$ then we define
\be\label{eq:h(w)-def}
h(w): = h(x_1) \otimes h(x_2) \otimes h(x_3) \otimes h(x_4) \otimes h(x_5) \otimes h(x_6) \in {\rm End}( (\IC^2)^{\otimes 6} )
\ee

\bigskip
\noindent
\textbf{Important Remark:}
An important point throughout this paper is that while the group $SU(2)^6$ acts
on the space of six Qbits, it does not act effectively. There is a subgroup
\be\label{eq:Z-def}
Z \subset Z(SU(2)^6) ~ ,
\ee
where $Z(SU(2)^6) \cong \IZ_2^6$ is the center,
that acts ineffectively on six Qbits. It is the subgroup of six signs whose product is $+1$. The group $Z$ is isomorphic to $\IZ_2^5$.
We will denote the quotient by $SU(2)_6 := SU(2)^6/Z$. (Some readers will prefer to use the notation $SU(2)^{\circ 6}$.)  Similarly,
we will denote the embedding of six copies of the quaternion group \eqref{eq:quatgroup} into $SU(2)^6$ by $Q^6$ and write $Q_6 := Q^6/Z$ for the quotient by $Z$. (Again, some readers will prefer to write $Q^{\circ 6}$.)
\bigskip

We may consider $h(w)$ to lie in the group $Q_6$. There is a non-split sequence
\be
1 \rightarrow \IZ_2  \rightarrow Q_6 ~~ {\buildrel \pi \over  \rightarrow} ~~ ( \IF_4^+)^6 \rightarrow 1
\ee
and $h$ defines a section of $\pi$ with cocycle
\be
h(w_1) h(w_2) = \chi(w_1, w_2) h(w_1+w_2)
\ee
given by the \underline{product} of the cocycles \eqref{eq:h-cocycle-def}.
\be
\chi(w_1,w_2)= \prod_{\alpha=1}^6 \epsilon(x_\alpha, y_\alpha)\, .
\ee
Note that this is to be distinguished from the cocycles generated by
the group elements $(h(x_1), \dots, h(x_6)) \in Q^6$. In the latter case, the
cocycle is the $6$-tuple of cocycles defined in  \eqref{eq:h-cocycle-def}.
This distinction will be of crucial importance in section  \ref{subsec:44Preserving}
below.

We now come to a central claim for this paper:
When restricted to the hexacode the cocycle defined by $h(w) \in Q_6$ is exactly equal to one.
That is, $h$ in fact defines a group homomorphism:
\be\label{eq:Hexacode-Split}
h(w_1) h(w_2) = h(w_1+w_2)\qquad\qquad w_1, w_2 \in \CH_6 \, .
\ee
This fact is both surprising and significant.
To prove it we write two hexacode words  $w_1, w_2$ as:
\be
\begin{split}
w_1 & = (x_1, x_2, x_3, x_1 + x_2 + x_3, x_1 \uwb + x_2 \uw + x_3 , x_1 \uw + x_2 \uwb + x_3) \, , \\
w_2 & = (y_1, y_2, y_3, y_1 + y_2 + y_3, y_1 \uwb + y_2 \uw + y_3 , y_1 \uw + y_2 \uwb + y_3) \, . \\
\end{split}
\ee
Now we use the bimultiplicative property to expand out
\be
\epsilon(x_1+x_2 + x_3 , y_1 + y_2 + y_3) = \prod_{i,j=1}^3 \epsilon(x_i,y_j)
\ee
and similarly for the other two nontrivial factors. Now, gathering all the terms together
we group together terms of the form  $\epsilon( a x_i, b y_j)$ for each pair $(i,j)$.
For the terms with $i=j$ we have in all
\be
\begin{split}
\epsilon(x_1,y_1) \epsilon(x_1,y_1)\epsilon(\uwb x_1, \uwb y_1) \epsilon(\uw x_1, \uw y_1) \\
\epsilon(x_2,y_2) \epsilon(x_2,y_2) \epsilon(\uw x_2, \uw y_2)\epsilon(\uwb x_2, \uwb y_2) \\
\epsilon(x_3,y_3) \epsilon(x_3,y_3) \epsilon(x_3, y_3)\epsilon(\uwb x_3, \uwb y_3) \\
\end{split}
\ee
and using the permutation property we see these all multiply to $1$. As an example of
$i\not=j$ consider
\be
\begin{split}
\epsilon(x_1,y_2) \epsilon(\uwb x_1, \uw y_2) \epsilon(\uw x_1 , \uwb y_2) & =
\epsilon(x_1,y_2) \epsilon(  x_1, \uwb y_2) \epsilon(  x_1 , \uw  y_2) \\
& =
\epsilon(x_1, (1+\uw + \uwb ) y_2)\\
&  = \epsilon(x_1, 0 \cdot y_2)\\
&  = 1\\
\end{split}
\ee
and it is similar for the other pairs.

It follows from \eqref{eq:Hexacode-Split} that
\be\label{eq:HexacodeProjector}
P = \frac{1}{64} \sum_{w \in \CH_6} h(w)
\ee
is a projection operator. Note that for $w\not=0$ we have $\Tr(h(w))=0$ and therefore
\be
\Tr(P) = 1
\ee
Therefore $P$ is a rank one projection operator. We will show in Section
\ref{subsec:ImP-SC} that for a suitably normalized spinor $s\in \Im P$, $V_s$ is a
superconformal current.

As a check we note that indeed $\Psi$ can in fact be written as:
\be\label{eq:Psi-hexacode}
\Psi = \frac{1}{4} \sum_{w\in \CH} h(w) \vert 0^6 \rangle
\ee

One way to check \eqref{eq:Psi-hexacode} is to use the
homomorphism $\pi$ defined in equation \eqref{eq:pi-projection} and its extension to $(\IF_4^{+})^6$.
Recall that this homomorphism is distinguished because it is
zero or one according to whether the section $h(x)$ is diagonal or off-diagonal, respectively.
So it is zero or one according to whether $h(x)$ induces a bit-flip.
One now checks that:
\be
h(x) \vert 0^6 \rangle = \chi(x) \vert \pi(x) \rangle  \qquad \qquad x \in \IF_4^6
\ee
where
\be
\chi(x) = 1^{\# \uz's} (-1)^{\# \uone's} (\I)^{\# \uw's} (-\I)^{\# \uwb's}
\ee

We can now prove \eqref{eq:Psi-hexacode} by noting that  the group of automorphisms of the hexacode
denoted by $H_0\times H_3$  in Appendix \ref{app:HexacodeAutGroup} induce corresponding symmetries of $\Psi$.
Using these symmetries
it suffices to check that we get suitable vectors from the seed codewords \eqref{eq:SeedWords}.
Then note that, restricted to the hexacode the fiber above $0^6$ is order four:
\be
\pi^{-1}(0^6) = \{ \uz\uz~ \uz\uz~ \uz\uz, \uz\uz~ \uwb\uwb~ \uwb\uwb, \uwb\uwb ~ \uz\uz ~ \uwb \uwb , \uwb \uwb ~ \uwb \uwb~ \uz\uz \}
\ee
So the sixtyfour hexacode words map to  sixteen different terms. As far as the phases are concerned it appears one just has to check these by hand.

Finally, we note that the image $\pi(\CH_6) \subset \IF_2^6$ is exactly the
 code of type $[6,4]$ described in section \ref{subsec:RelatedHamming} above: As promised
that code determines the $N=1$ superconformal generator.

\bigskip
\textbf{Remark}: The one-dimensional code defined by $\Psi$ is closely related to the standard
code of type $[[6,0,4]]$ discussed near equation \eqref{eq:StabCodeword}. Mainiero computed its
entanglement homologies and found them to be identical. Indeed, there is a local transformation
relating one to the other.  Mainiero found that
\be
\Psi_{[[6,0,4]] } = \left( 1 \otimes U_1 \otimes 1 \otimes U_2 \otimes U_2 \otimes 1 \right) \Psi
\ee
where
\be
\begin{split}
U_1  = \frac{1}{\sqrt{2}} \begin{pmatrix} 1& 1 \\ - \I & \I \\ \end{pmatrix} \qquad
& \qquad
U_2  = \frac{1}{\sqrt{2}} \begin{pmatrix} 1& - 1 \\ \I & \I \\ \end{pmatrix} \\
\end{split}
\ee

\subsection{Why The Image Of $P$ Defines A Superconformal Current} \label{subsec:ImP-SC}

We now show that, for a suitably normalized state $s\in (\IC^2)^{\otimes 6}$ in the image of $P$
the vertex operator $V_s$ is a superconformal current with $c=6$.

Given any two states $s^1,s^2 \in (\IC^2)^{\otimes 6}$ we have the operator product expansion
\be
V_{s^1}(z_1) V_{s^2}(z_2) \sim
\frac{\bar s^1 s^2}{z_{12}^3} + \kappa_1 \sum_{A} \frac{ \bar s^1 \Sigma^A s^2}{z_{12}^2} J^A  +
\kappa_1^2  \sum_{\alpha < \beta} \frac{ \bar s^1 \Sigma^{AB} s^2}{z_{12}}  J^A J^B +
\kappa_2 \frac{\bar s^1 s^2}{z_{12}}T+   + \cdots
\ee
where $J^A$ correspond to the 18 generators of $SU(2)^6$, $T$ is the energy-momentum
tensor and all operators on the RHS are evaluated at $z_2$. We use a composite index
$A=(\alpha,i)$. Here
$1\leq i \leq 3$ label a basis of three generators of $SU(2)$
and $1\leq \alpha \leq 6$.   (Later we will interpret $\alpha$  as the column number in the Miracle
Octad Generator.)  Then $\Sigma^A$ is a
Pauli matrix for the column $\alpha$:
\be
\Sigma^A = 1 \otimes \cdots  \otimes (\sigma^i)^{(\alpha)} \otimes \cdots 1
\ee
The conjugate spinor is defined by
\be
\bar s :=  s^{tr} \bar \gamma
\ee
where
\be
\bar \gamma = (\I \sigma^2)^{(1)} \otimes \cdots \otimes (\I \sigma^2)^{(6)}
\ee
is a symmetric matrix.
The constant $\kappa_1$ depends on our normalization of $SU(2)$ currents. With the convention
(again dropping cocycle factors)
\be
\begin{split}
J^\pm & \sim e^{\pm\I \sqrt{2} X} \\
J^3 & = - \I \p X \\
\end{split}
\ee
We find  $\kappa_1 = 1/\sqrt{2}$. Specializing to $s^1 = s^2 = s$ and using the symmetry properties
of the Pauli matrices so that $\bar s \Sigma^A s =0$ the above equation simplifies to
\be
V_{s}(z_1) V_{s}(z_2) \sim
\frac{\bar s s}{z_{12}^3} +
\kappa_1^2  \sum_{\alpha < \beta} \frac{ \bar s \Sigma^{AB} s}{z_{12}}  J^A J^B +
\half \frac{\bar s s}{z_{12}}T +   + \cdots
\ee
Comparison with line one of equation \eqref{eq:N1-SCA-OPE} shows that the
conditions for $V_s$ to be a superconformal current are that
\be\label{eq:SimplerN1}
\begin{split}
\bar s \Sigma^{AB} s & = 0  \qquad 1 \leq \alpha < \beta \leq 6 \\
\end{split}
\ee
once these equations are satisfied one can normalize $s$ to achieve the
desired OPE of the supercurrent. In the present case $c=6$ corresponds to
$\bar s s = 1$.

The quadratic equations in \eqref{eq:SimplerN1} are, as far as we are aware,  all independent
so there are $ {6 \choose 2} \times 9 = 135$ independent equations on $64$ variables.
Nevertheless, if $s \in \Im P$ then indeed the equations are satisfied. To see this,
we define a real structure on $(\IC^2)^{\otimes 6}$ so that the basis $[abc\dots]$ are real
vectors. Then
\be
\bar \Psi = \I \Psi^\dagger ,
\ee
so $\bar \Psi \Sigma^{AB}\Psi=0 $ iff $\Psi^{\dagger} \Sigma^{AB} \Psi =0$. We will now
proceed to check that these equations indeed hold.
Since $\Sigma^{AB}$ represent bit-flip and/or phase-flip errors on two Qbits,  this is
indeed the quantum error-detecting property of the code! Note that this is really a property
of the {\it quantum} code, and not the classical code. For example.
\be
\begin{split}
\Sigma^{(1,1), (2,1)} \Psi & = [12] + \I[3456] + ([34] + [123456] + [56]) \\
& + \I ([\emptyset] + [1256] + [1234] ) + ([235] + [145]+[136] + [246] ) \\
& - \I ( [146] + [135] + [236] + [245] )  \\
\end{split}
\ee
\be
\begin{split}
\Sigma^{(1,1), (2,2)} \Psi & = - [12] + \I[3456] + ([34] - [123456] + [56]) \\
& + \I ([\emptyset] - [1256] - [1234] ) + (-[235] + [145]+[136] - [246] ) \\
& + \I ( - [146] + [135] - [236] - [245] )  \\
\end{split}
\ee
All the classical code words are the same as in $\Psi$, but, remarkably, these vectors
are in fact orthogonal to $\Psi$.

The symmetry properties of the hexacode mentioned near \eqref{eq:SeedWords} imply
similar symmetry properties of $\Psi$ that
allow us to map the general case of $\Sigma^{AB}$ to two cases:   first, if $\alpha, \beta$ are
in the same couple then by permuting within couples and using the cyclic symmetry of the 3 Pauli matrices
we reduce to the two cases above. Second, if  $\alpha,\beta$ are in different couples then
we need to check orthogonality to
\be
\Sigma^{(1,i),(3,j)}
\ee
and then we need only check $(i,j)=(1,1)$ and $(i,j) = (1,2)$. But in this case both $i,j$ involve bit-flips
and one easily checks that a pair of bit-flips on $1,3$ maps every term of $\Psi$ to a vector orthogonal to
every term in $\Psi$. So, rather trivially, $\Sigma^{AB} \Psi$ is orthogonal to $\Psi$ for these cases.
Thus $\bar \Psi \Sigma^{AB} \Psi =0$ for all $AB$ with $\alpha \not= \beta$. Finally, since
$\bar \Psi \Psi = 16 \I$, we conclude that $V_s$ with $s = \frac{e^{-\I \pi/4}}{4} \Psi$ defines
a superconformal current with $c=6$.

\subsection{The Relation Of The $N=4$ Supercurrents To Codes}\label{subsec:N=4AndCodes}

Now that we have understood the code underlying $N=1$ superconformal
symmetry we can easily describe the $N=4$ supercurrents in terms of
the hexacode.

For each $x\in \IF_4$ let $\CH_6^x$ denote the set of hexacode words
whose first digit is $x$, and let
\be
P_x:= \frac{1}{16} \sum_{w\in \CH_6^x} h(w)
\ee
Note that $\CH_6^0$ is a linear subspace of $\CH_6$. In fact, it has
$16$ elements and is therefore a two-dimensional subspace. One choice of
basis is
\be
u_1 = (\uz,\uz, \uz,\uz,\uone, \uone) \qquad  u_2 = (\uz, \uwb, \uwb, \uz, \uw, \uone) \, .
\ee
The $\CH_6^x$ are the cosets of $\CH_6^0$ in $\CH_6$ and therefore
each has sixteen elements. It follows that
\be
P_x P_y = P_{x+y}
\ee

Now, $P_{\uz}$ is a projection operator and its image is $4$-dimensional.
The first  $SU(2)$ factor in $SU(2)^6$ will be interpreted as the
$R$-symmetry group. Note that it commutes with $P_{\uz}$
and therefore $\Im(P_{\uz})$ is an $SU(2)_R$ representation. Focusing on the first
Qbit, $\vert \pi(x) \rangle$ span a two-dimensional space so the
representation is $\textbf{2} \oplus \textbf{2}$.
The four supercurrents will span the image of $P_{\uz}$.
In fact, we have
\be
\begin{split}
\Psi^1 & = P_{\uz} (1 + P_{\uwb}) \vert +^6 \rangle \\
\Psi^2 & =  P_{\uz} (-1 + P_{\uwb}) \vert +^6 \rangle \\
\bar \Psi^1 & = P_{\uz}( P_{\uone} + P_{\uw}) \vert +^6\rangle \\
\bar \Psi^2 & = P_{\uz} (-P_{\uone} + P_{\uw} ) \vert +^6\rangle \\
\end{split}
\ee
(Note that $\Psi^2 = \exp[-\I \pi T^2] \Psi^1 = h(\uone)^{(1)} \Psi^1$,
and that $h(\uone)$ anticommutes with $h(\uw)$ and $h(\uwb)$.)

\textbf{Remarks}:

\begin{enumerate}

\item The image $\pi(\CH_6^0) \subset \IF_2^{6}$ is the span of
the vectors $w_1, w_2, w_3$ in the $[6,3]$ subcode of the truncated
Hamming code, as described in section \ref{subsec:RelatedHamming}.

\item For any nonzero $x\in \IF_4$, $\CH_6^0 + \CH_6^x$ is closed
under vector addition and therefore $\half(P_{\uz} + P_x)$ is a
projector to a two-dimensional subspace of $\Im(P_{\uz})$. This defines
an embedding of an $N=2$ subalgebra in the $N=4$ superconformal algebra.

\end{enumerate}

\section{Relation To Mathieu Moonshine}\label{sec:Moonshine}

\subsection{Statement Of Mathieu Moonshine}\label{sec:MathaiMoonSumm}

The RR subspace of the space of states of a K3 sigma model is a representation
of the $(4,4)$ superconformal algebra and admits an isotypical decomposition
\bel
\CH_{RR} = \oplus_{i,j}    D_{i;j}  \otimes R_i \otimes \tilde R_j
\ee
where $R_i$  runs over unitary irreps of leftmoving N=4 and  $ \tilde R_j $ runs over  unitary irreps of rightmoving $N=4$.
In particular  $i$ or $j$ corresponds to a pair $(h,\ell)$ with $h\geq 1/4$, $\ell = 1/2$ for $h>1/4$ and
$\ell \in \{ 0,1/2 \}$ when $h=  1/4$.

The remarkable Mathieu Moonshine conjecture \cite{Eguchi:2010ej}
is that the virtual
\footnote{On the Kummer locus the spaces $D_{i;\tilde h=1/4, \tilde \ell =1/2}$
with $h_i>1/4$ are nonzero but there is a general expectation that these spaces vanish off the Kummer locus.}
degeneracy space
\be\label{eq:DegeneracySpaces}
\CD_i := D_{i; \tilde h=1/4, \tilde \ell=0} - 2 D_{i;\tilde h=1/4, \tilde \ell =1/2}
\ee
for $i$ with $h>1/4 $ is, in some natural (but thus far unexplained) way,
a representation of the finite simple group $M_{24}$. Moreover, these $M_{24}$ representations have the property
that the character-valued extension of the elliptic genus exhibits modular properties.
This is remarkable because, thanks to the quantum Mukai theorem described below,
$M_{24}$ is not the quotient of a group of (4,4)-preserving automorphisms of
any single K3 sigma model. The significance of the virtual degeneracy space
\eqref{eq:DegeneracySpaces} arises from consideration of the elliptic genus. The Witten index of
R-sector irreducible representations of the N=4 superconformal algebra is:
\be
{\rm Tr}_{R_{(h,\ell)} }e^{ 2\pi \I J_0^3 } q^{L_0-c/24} = \begin{cases}
1 & h=1/4, \ell= 0 \\
-2 & h=1/4 , \ell = 1/2 \\
0 & h>1/4, \ell = 1/2\\
\end{cases}
\ee
Therefore, if $g \in \Aut(\CC)$ is any finite-order automorphism of a K3 sigma model $\CC$ that
commutes with the $(4,4)$ superconformal algebra then the
twisted character
\be
\begin{split}
\CE_g(z,\tau)& := {\rm Tr}_{\CH_{RR}} g e^{2 \pi \I (J_0^3 + \tilde J_0^3)} q^{L_0 - c/24} e^{2\pi \I z J_0^3} \bar q^{\tilde L_0 - c/24}  \\
& = \sum_i  \Tr_{\CD_i}(g) \chi_i(\tau,z) \\
\end{split}
\ee
must be a weight zero index one Jacobi form for a congruence subgroup of index determined by the order of $g$.
Here $\chi_i(\tau,z)$ are characters of irreducible representations of the (left-moving) $N=4$ algebra:
\be
\chi_{h,\ell}(z,\tau)
:= {\rm Tr}_{R_{(h,\ell)} }e^{ 2\pi \I J_0^3 }e^{ 2\pi \I z J_0^3 } q^{L_0-c/24}
\ee

The outcome of further investigations \cite{Cheng:2010pq, Gaberdiel:2010ca, Eguchi:2010fg,Gannon}
is that there exist an infinite set of representations of the group $M_{24}$:
\be\label{eq:H-type-reps}
H_{0,0}, H_{0,1/2}, H_n \qquad n\geq 1
\ee
(where $H_{0,0} = \textbf{23} - 3 \cdot \textbf{1}$ and $H_{0,1/2} = - 2\cdot \textbf{1}$, but all the other $H_n$ are
true, not virtual, representations)
such that for {\it every} $g\in M_{24}$ the function
\be\label{eq:PseudoEG}
\hat \phi_g(z,\tau) :=  \Tr_{H_{0,0}}(g) \chi_{h=1/4, \ell=0} + \Tr_{H_{0,1/2}}(g) \chi_{h=1/4,\ell=1/2}
+ \sum_{n=1}^\infty (\Tr_{H_n}(g)) \chi_{n+1/4,\ell=1/2}
\ee
transforms  {\it as if} $g \in M_{24}$ acted on
$\CC$ as a $(4,4)$-preserving automorphism.  Moreover
\be\label{eq:Equality}
\hat \phi_g(z,\tau) = \CE_g(z,\tau)
\ee
in those cases where  $g$ is truly a $(4,4)$-preserving automorphism of the CFT. However, we stress again that there is no known
natural action of $M_{24}$ on the spaces $\CD_i$.
We note however that in \cite{Taormina:2013mda} A. Taormina and K. Wendland discuss how a certain
maximal subgroup of $M_{24}$, the octad group, acts on a $45 + 45^*$ dimensional subspace
of the states of $(h,\bar h) = (5/4, 1/4)$. Some extensions of this work appear in \cite{Gaberdiel:2016iyz,Keller:2019suk}.

The most natural way to explain the Mathieu Moonshine phenomenon would
be to find {\it some} K3 sigma model with a $(4,4)$-preserving
automorphism group that has a quotient that contains $M_{24}$.
\footnote{We stress that all that is needed is that some quotient of
the automorphism group contains $M_{24}$ as a subgroup. The action
of the $(4,4)$-preserving automorphism group $G$ on any given
degeneracy space $D_{i;j}$ might, in general, have a kernel. Note that a different
quotient, not containing $M_{24}$ as a subgroup, could act on the
massless states where $h_i=1/4$, thus explaining why these are not
true $M_{24}$ representations. }
However, the quantum Mukai theorem of Gaberdiel, Hohenegger, and Volpato,
reviewed in section \ref{sec:Mukaif} below is a powerful no-go theorem
that implies that such an explanation of Mathieu Moonshine cannot work.
Thus, one must relax some of the hypotheses of the quantum Mukai theorem.

\subsection{The Mukai Theorem}

When a K3 surface is given a complex structure it is holomorphic symplectic.
The Mukai theorem characterizes the possible groups of holomorphic symplectic
automorphisms of K3 surfaces. For a nice review see \cite{MasonK3}. Once a
K3 surface is endowed with a complex structure the 24-dimensional cohomology
space has a Hodge decomposition
\be
H^*(K3;\IC) \cong H^{0,0} \oplus H^{2,0} \oplus H^{1,1} \oplus H^{0,2} \oplus H^{2,2}
\ee
Any holomorphic automorphism must preserve these five components. The Mukai
theorem says that all groups of holomorphic symplectic automorphisms are subgroups
of $M_{23}$ with at least $5$ orbits in the natural action of $M_{23}$ on a set
of 24 elements. (The group $M_{24}$ has a natural action as a permutation subgroup
acting on a set with $24$ elements.
The subgroup $M_{23}$ is isomorphic to any subgroup preserving one element.)

\subsection{Quantum Mukai Theorem} \label{sec:Mukaif}

Motivated by the discovery of Mathieu Moonshine, Gaberdiel, Hohenegger, and Volpato (GHV)
gave a characterization of the potential automorphism groups of supersymmetric
K3 sigma models that preserve $(4,4)$ supersymmetry. The answer is, remarkably,
that the groups are precisely the subgroups of the Conway group that preserve sublattices
of the Leech lattice of rank greater than or equal to four \cite{Gaberdiel:2011fg}.
In order to prove the theorem one follows \cite{Aspinwall:1994rg,Aspinwall:1996mn}  to characterize
a K3 sigma model as a choice of positive definite four-dimensional subspace of the
Grassmannian $O(4,20)/O(4) \times O(20)$. A clever argument transfers the problem from a question about spaces of indefinite signature $(4,20)$
to questions about actions on the Leech lattice, a positive definite lattice of rank $24$.
(The proof is elegantly summarized in \cite{Huybrechts:2013iwa}. See also  \cite{Taormina:2011rr}
for related remarks.)

Now the subgroups of the Conway group that preserve sublattices of the Leech lattice have
been tabulated in \cite{HM}. None of the relevant groups contains $M_{24}$ as a subgroup of
a quotient. (Also, although this is less relevant,   many groups cannot be embedded as subgroups of $M_{24}$.)
 The Quantum Mukai Theorem is thus a powerful no-go statement in the search for an explanation of Mathieu Moonshine.

Among the rank four H\"ohn-Mason groups there is
a distinguished maximal subgroup,
$2^8{\,:\,}M_{20}$. This is closely related to the GTVW model \cite{gtvw}.
Clearly such a special point deserves special attention.
Using the relation of supercurrents to codes we will give a
different derivation of the main result of \cite{gtvw}.
Our approach makes it clear that a maximal subgroup of the
Mathieu group, namely the sextet group, acts as a group of
$(1,1)$-preserving automorphisms of the model.

\subsection{Evading A No-Go Theorem}

Given the powerful no-go theorem of GHV, any explanation of
Mathieu Moonshine must proceed by relaxing one of the hypotheses in
the theorem.

Two ways of relaxing the hypotheses  have been explored in the
past, but, at least thus far,  have only met with partial success. One approach is to posit
 that $M_{24}$ acts not as an automorphism of the full conformal field theory but only
 as an ``automorphism of the subspace of BPS states.'' The BPS states are the $N=4$ primaries
 with left-moving quantum numbers $(h=1/4 + n , \ell = 1/2)$ and right-moving
 quantum numbers $(\tilde h = 1/4, \ell=0,1/2)$.
 To make sense of this idea one would need some ope-like algebra structure
on these BPS states. We could call this the ``algebra of BPS states'' approach following \cite{Harvey:1995fq, Harvey:1996gc}.
There has been some success with this approach in the context of moonshine
\cite{Paquette:2017xui}, but not yet in the context of Mathieu moonshine.
A second approach is to attempt to ``combine'' the symmetries of different
K3 sigma models at different points in the moduli space. This is the
``symmetry-surfing'' approach that has been vigorously pursued by A. Taormina
and K. Wendland \cite{Taormina:2011rr, Taormina:2013jza}.

The project described in this paper began with the observation that one could
 relax the assumption that the relevant group of automorphisms
of the K3 sigma model commutes with $(4,4)$ supersymmetry. Thus, we are
imagining that the $g \in M_{24}$ which can be used to define $\hat \phi_g(\tau,z)$
are in fact true automorphisms of at least some K3 sigma model but do not commute with
$(4,4)$ supersymmetry. In order for the Witten index to make sense, $g$ must still commute
with some right-moving $N=1$ supersymmetry. We are thus led to the idea that
there might be very symmetric K3 sigma models with large symmetry groups that
commute with $(4,1)$ supersymmetry, and that these symmetry groups contain
$M_{24}$ as a quotient group. While this idea was an important motivation for
our work, we will argue in section \ref{subsec:NewTwined} below, using the twined elliptic genera,
that, at least for the GTVW model, the enhancement from $(4,4)$ preserving to $(4,1)$ preserving
symmetries will not explain $M_{24}$ symmetry. So the mystery of Mathieu Moonshine remains.

The statement of Mathieu Moonshine is only slightly altered in the $(4,1)$ case.
The irreducible representations of the $N=1$ superconformal group are labeled
with $(h,\epsilon)$ where $\epsilon$ is a sign given by the action of $(-1)^F$
on the groundstate and $h\geq c/24$ for unitarity. Only the representations
with $h=1/4$ have nonvanishing Witten index. Working out the
branching rules for $N=4$ representations in terms of $N=1$ representations
the only new point is that the virtual degeneracy spaces relevant to the elliptic genus are now
\be
\CD_i^{(4,1)}  := D_{i; h=1/4,+}^{(4,1)} - 4 D_{i;h=1/4, -}^{(4,1)}
\ee

\textbf{Remark}:
 In the spirit of looking for larger automorphism groups by reducing the
 amount of preserved supersymmetry it is natural to ask if one could
 consider instead the automorphisms of K3 sigma models that commute
 with $(2,1)$ supersymmetry.  This would indeed be possible
 if  all the representations in \eqref{eq:H-type-reps} were true $M_{24}$ representations.
 Unfortunately, because the massless ones are virtual, and the branching of massless N=4 reps to
  N=2 reps contains infinitely many massive N=2 reps, the virtual representations make infinitely many
  massive representations into virtual representations of $M_{24}$. Once one admits infinite
  numbers of virtual representations Moonshine becomes unsurprising.

\section{Symmetries Of Supercurrents}

The construction of supercurrents from quantum error correcting codes provides new insight into the subgroup of the symmetry group
that stabilizes the supercurrents. We are mainly interested in the group preserving certain superconformal structures in the GTVW model
but will also comment on the SCFTs with moonshine for the Conway group.

\subsection{The Stabilizer Group Within $Q^6$}\label{subsubsec:StabInQ6}

As a first step to determining the symmetries of the supercurrent we note
that it follows from eqn. \ref{eq:Hexacode-Split} and the expression \ref{eq:Psi-hexacode} for $\Psi$ that
$h(w) \Psi = \Psi$ for all $w \in \CH_6$.
Thus a copy of $\CH_6 \subset Q_6$ is a group of symmetries of the supercurrent.

Now, recalling the definition \eqref{eq:Z-def} we would like to lift this to the
group $Q^6 \subset SU(2)^6$ of bit-flip and phase-flip errors. Recall that
\be
1 \rightarrow Z \rightarrow Q^6 \rightarrow Q_6 \rightarrow 1
\ee
We now determine the stabilizer group ${\rm Stab}_{Q^6}(\Psi)$.
We claim this is the non-abelian group
\be\label{eq:Stab-Psi-Q6}
{\rm Stab}_{Q^6} (\Psi) = \{\left(
 \epsilon^1 h(x_1) , \dots ,   \epsilon^6 h(x_6) \right) \vert \prod_{\alpha=1}^6 \epsilon^\alpha = 1 \, \, {\rm and} \, \,(x_1, \dots, x_6 ) \in \CH_6 \}
\ee

To prove this, we note that the elements of $Q^6$ can be written as $(\epsilon^\alpha h(x_\alpha))_{\alpha=1}^6$.
Let $w=(x_1, \dots, x_6) \in \IF_4^6$. Then we need to solve
\be
\prod_{\alpha=1}^6 \epsilon^\alpha  h(w) \Psi = \Psi
\ee

Now, note that   $h(w)$  for $w\in \IF_4^6$  form a linear basis for $\End( (\IC^2)^{\otimes 6})$.
This follows since   $h(x)$ for   $x\in \IF_4$ form a linear basis for the complex vector space $\End(\IC^2)$.
Therefore  $h(w) \Psi$ for $w\in \CH_6^{\rm comp}$ form a linear basis for $(\IC^2)^{\otimes 6}$.
This follows because the vectors $h(w) \Psi$ for $w\in \IF_4^6$ generates the entire vector space.
But every $w\in \IF_4^6$ can be written as $w= u_1 + u_2$ with $u_1 \in \CH_6^{\rm comp}$
and $u_2 \in \CH_6$ (as we saw in equation \eqref{eq:MaxIsotropicDecomp} et. seq. above)
and then $h(w)\Psi = \pm h(u_1) \Psi$. So these must generate the
entire vector space. But there are at most $4^3 = 2^6$ linearly independent vectors
$h(u_1)\Psi$, so these must in fact be linearly independent. It follows that if
there is a $w\in \IF_4^6$ such that $h(w)\Psi = \pm \Psi$ then the sign must be $+$ and  $w\in \CH_6$.

Thus we have
\be\label{eq:LiftCH6}
1 \rightarrow Z \rightarrow {\rm Stab}_{Q^6} (\Psi) \rightarrow \CH_6 \rightarrow 0
\ee
where the reader will recall that $Z$ is the subgroup of the center of $SU(2)^6$ that
acts ineffectively on the 6 Qbit system.

\subsection{Further Symmetries Of $\Psi$: Lifting The Hexacode Automorphisms}\label{subsec:HexAutLift}

The group $H:=SU(2)^6 : S_6$ acts on the six Qbit system in a natural way,
and this group can be lifted to a symmetry group of the chiral part of the GTVW model.
In sections \ref{subsec:44Preserving} and \ref{subsec:41Preserving}
below we discuss the lift to the full GTVW model. As a preliminary, it is therefore useful to discuss
what we know about ${\rm Stab}_{H}(\Psi)$. Here it is very useful to observe that
automorphisms of the hexacode lift to operators on the Q-bit system that commute
with the projector $P$ defined in \eqref{eq:HexacodeProjector}.

To demonstrate this we use the description of the automorphism group of the hexacode
in Appendix \ref{app:HexacodeAutGroup}. The generators $g_1, \dots, g_4$ are pure permutations.
Letting $\hat g_1, \dots, \hat g_4$ denote the corresponding permutations acting on
the factors of the six Qbit Hilbert space we clearly have
\be
\hat g_i h(w) \hat g_i^{-1}  = h(g_i\cdot w)
\ee
for all $w\in \IF_4^6$ and all $i=1,2,3,4$ and therefore $\hat g_i$ commutes with  $P$.
For $g_0$ we note that $h(\omega x) = \Omega^{-1} h(x) \Omega$ and therefore
letting $\hat g_0: = (\Omega^{-1})^{\otimes 6}$ we have
\be
\hat g_0 h(w) \hat g_0^{-1} = h(g_0\cdot w)
\ee
for all $w\in \IF_4^6$ and therefore $\hat g_0$ commutes with $P$.
Similarly, define
\be\label{eq:hatg5-def}
\hat g_5:= (465) \cdot (\Omega^{-1} \otimes \Omega \otimes 1 \otimes 1 \otimes 1 \otimes 1)
\ee
so that
\be
\hat g_5 h(w) \hat g_5^{-1} = h(g_5 \cdot w)
\ee
for all $w\in \IF_4^6$. Accordingly, $\hat g_5$ commutes with $P$.

Finally, we define a lift of $g_F$. It is easy to prove that there is
no linear operator that implements the Frobenius automorphism on $\IF_4$.
That is, there is no matrix $A$ such that  $A h(x) A^{-1} = h(\bar x)$. Nevertheless, if
we define
\be
v:= \frac{\I}{\sqrt{2}}(\sigma^1 + \sigma^3) = \frac{\I}{\sqrt{2}}\begin{pmatrix}  1 & 1 \\   1&  - 1 \\ \end{pmatrix}
\ee
one can check that $v^2=-1$ and
\be
v h(x) v^{-1} = h(x)^\dagger =  \begin{cases}
h(x) & x= \uz \\
-h(\bar x) & x\not=\uz \\
\end{cases}
\ee
 Define $\hat g_F = (56) \cdot v^{\otimes 6}$. Then,
since there are always an even number of nonzero digits in a hexacode word we do in fact have
\be
\hat g_F h(w) \hat g_F^{-1} = h(g_F \cdot w)
\ee
when $w\in \CH_6$, and this is sufficient to prove that $\hat g_F$ commutes with $P$.

Thus, the lifted elements $\hat g_0, \dots,\hat g_5,  \hat g_F$ generate a group that commutes with
$P$.
Therefore, since the image of $P$ is one-dimensional we can say that
\be
\hat g_i \cdot \Psi = \xi_i \Psi  \qquad i=1, \dots , 7
\ee
for some phase $\xi_i$. Now, for $\hat g_1, \dots, \hat g_4$, a simple direct check shows that
$\xi_i = +1 $ for $i=1,2,3,4$. We must work a bit harder to find $\xi_0$, $\xi_5$,  and $\xi_F$.

Since $\hat g_0^3 = 1$ it follows that $\xi_0$ is a third
root of unity.  We claim that, in fact, $\xi_0 =1$. To prove this we use the reality properties
of $\Psi$. Define the symmetric matrix:
\be\label{eq:Def-bar-gamma}
\bar \gamma = (\I \sigma^2)^{(1)} \otimes \cdots \otimes (\I \sigma^2)^{(6)}
\ee
and compute:
\be\label{eq:Psi-cplxconj}
\begin{split}
\bar \gamma \Psi & = \I \Psi^*   ~ .  \\
\end{split}
\ee
Now recall that    $  \Omega^*  = -  (\I \sigma^2) \Omega (\I \sigma^2) $ which implies that  $\hat g_0^* = \bar\gamma \hat g_0 \bar \gamma$
and then  \eqref{eq:Psi-cplxconj} implies that  $\xi_0^* = \xi_0$  and therefore $\xi_0=1$. The same
style of argument shows that $\xi_5 = +1$.

It is worth noting that since $\hat g_i\Psi=\Psi$ for $i=0,\dots, 5$ it follows that
the group $\langle \hat g_0, \dots, \hat g_5 \rangle$, which is, \emph{a priori}, only
an extension of $H_{05}:=\langle g_0, \dots, g_5 \rangle $ is in fact isomorphic to $H_{05}$.

Finally, we note that $\hat g_F$ has order two. A direct computation
of $\langle +^6 \vert \hat g_F \vert \Psi\rangle$ shows that in fact
\be\label{eq:gF-acts-minus}
\hat g_F \Psi = - \Psi
\ee
so $\hat g_F$ is not in the stabilizer group. However we can remedy this by defining $\check g_F$ to be the product of $\hat g_F$
with the transformation $(-1,1^5) \in SU(2)^6$ (or any other element of the  center of $SU(2)^6$ that is not in $Z$)
\footnote{We thank T. Johnson-Freyd for pointing this out to us.}. Then
\be
\check g_F \Psi = \Psi
\ee

As discussed in Appendix \ref{app:HexacodeAutGroup} the elements $g_0, \dots, g_5,g_F$ generate $\Aut(\CH_6)$
 isomorphic to $\IZ_3 . S_6$.  The   lifts $\hat g_0, \dots, \hat g_5, \check g_F$ stabilize $\Psi$
 and generate a group isomorphic to $\IZ_3.S_6$.  Of course, the ``translation'' action by hexacode elements
themselves stabilize $\Psi$ and so the semidirect product
\be\label{eq:Holomorph-StabPsi}
2^6:3.S_6  \subset SU(2)_6 : S_6
\ee
stabilizes $\Psi$.

\textbf{Remarks}

\begin{enumerate}

\item
What we have described above as the stabilizer of $\Psi$ is the holomorph of the hexacode, ${\rm Hol}(\CH_6)$.
See Appendix \ref{app:HexacodeAutGroup} for a definition of the term ``holomorph.''

\item We still must lift the above symmetry group in $SU(2)_6 : S_6$
to a symmetry of the full GTVW Hilbert space. There are two issues involved
when doing this. First, lifting from a subgroup of $SU(2)_6$ to $SU(2)^6$
involves an extension by the subgroup $Z$ of the center of $SU(2)^6$.
Second, including left-movers with right-movers, those automorphisms that
involve a nontrivial permutation of hexacode digits must be embedded
diagonally in the product of left- and right-moving hexacode holomorphs.
These aspects will be carefully described in sections \ref{subsec:44Preserving}
and \ref{subsec:41Preserving} below.
\end{enumerate}

\subsection{The Stabilizer Group Of $Im(P)$ Within $SU(2)^6$ }\label{subsec:FiniteStab}

The question now arises as to the nature of the full stabilizer group
within the automorphism group $SU(2)^6:S_6$ of the chiral algebra $\CV_{\vec 0}$
of the GTVW model.

One nice consequence of the error-detecting code description of $\Psi$
is that the stabilizer within $SU(2)^6$ is a discrete  group. To show this
we consider the $h(x_i)$ for $x_i \in \IF_4^*$ to be generators
of the Lie {\it algebra} $\fs\fu(2)$. Note well that in this sub-section
we are not  thinking of these matrices multiplicatively!

We can show that the stabilizer group of $\Psi$ is discrete
by showing that there are no nontrivial solutions of
\be\label{eq:Inf-Stab}
\sum_{x\in \IF_4^*} \sum_{\alpha=1}^6  c_{x,\alpha} h(x)_\alpha \Psi =0
\ee
where $h(x)_\alpha$ means that the matrix only acts on the $\alpha^{th}$
factor. The computation, which is slightly technical, is relegated to
Appendix \ref{app:ProofDiscrete}. The main point though, is that it is a
again true due to the error correcting properties of $\Psi$.

In fact,  the stabilizer is a {\it finite}
group.  If the stabilizer were discrete and infinite, then, being a
subgroup of the compact group $SU(2)^6$ there would be an
accumulation point. We can rule out this possibility by noting that
in fact the stabilizer group is an algebraic group. Indeed, it can
be characterized as the solutions to
\be
\langle \Psi \vert u \vert \Psi \rangle = \langle \Psi \vert \Psi  \rangle
\ee
which constitutes a (complicated) algebraic equation for the matrix
elements of $u\in SU(2)^6:S_6$.

In \cite{Johnson-Freyd:2019wgb}   T.~Johnson-Freyd
has discussed the automorphism groups of holomorphic $N=1$ superVOA's
in a large class of models. It turns out that the GTVW model is a special
case of the class of models considered in \cite{Johnson-Freyd:2019wgb}.
Using the methods of \cite{Johnson-Freyd:2019wgb}, and the relation of the GTVW model
to a theory of $12$ MW fermions, one can show that the symmetry group of
$\Psi$ can in fact be no larger than the holomorph of the hexacode. Given
our result above, it is exactly the holomorph of the hexacode.

\subsection{Extending The Automorphism Groups To Include Left- and Right-Movers }

The $SU(2)$ $k=1$ WZW model has, famously,  $\widehat{SU(2)}^{k=1}_L \times\widehat{SU(2)}^{k=1}_R$ current
algebra symmetry and the global $SU(2)_L \times SU(2)_R$ symmetry is an automorphism of the model.
(An automorphism of the sigma model should certainly preserve the conformal weights, and hence we
only take the subgroup of the affine Lie group that commutes with $L_0$ and $\tilde L_0$.)
Actually, the diagonally embedded center of $SU(2)_L \times SU(2)_R$ acts trivially so only the
quotient, denoted $SO(4)_{LR}$,  acts effectively. On the other hand, there is a left-right reflection action
of the model. It exchanges the left- and right-movers and generates an $O(4)$ action on the space of states.
Elements in the nontrivial component permute conformal weights $(h,\tilde h) \to (\tilde h, h)$
and hence $O(4)$ is not an automorphism of the entire sigma model. Nevertheless, it is a  global
symmetry of the space of ground states and it can be useful.

When we turn to the product of six WZW models we clearly have a symmetry group
$SO(4)_{LR}^6 : S_6$, where the permutation group $S_6$ permutes the
$6$ factors. In the GTVW spectrum \eqref{eq:xGTVW-HS} there are spinor representations
so the symmetry group is in fact a quotient of ${\rm Spin}(4)^6: S_6$. Let $Z_{\rm diag}$
denote the diagonal embedding of  $Z \subset SU(2)^6$ into $\left( SU(2)_L^6 \times SU(2)_R^6 \right)$.
The group of symmetries acting effectively on the GTVW spectrum is
$\left({\rm Spin}(4)^6/Z_{\rm diag}\right): S_6$.

Just the way there is a further parity symmetry when one considers ground states of the
WZW model for a single factor $SU(2)$, there is   an $O(4)^6 : S_6$ symmetry group
of the set of RR ground states of the model.
In the GTVW model we find spinor representations and there is a group action of $\Pin(4)^6 : S_6$
where $\Pin(4)$ is the double cover of $O(4)$ that acts on spinors.
\footnote{Thanks to Bott periodicity $\Pin^+(4) \cong \Pin^-(4)$. In fact, the groups
are canonically isomorphic: If $e_i$ is a set of Clifford generators for $\Pin^\pm(4)$ then
$\frac{1}{3!} \epsilon_{ijkl} e_j e_k e_l $ is a set of Clifford generators for
$\Pin^{\mp}(4)$. Of course, they are not isomorphic as double covers of $O(4)$.  }
Note that ${\rm Pin}(4)^6$ has a projection to $O(4)^6$ and
taking the determinant of each factor gives a map to $\IZ_2^6$.
When the image is not one, some left- and right-conformal weights on some
factors will be exchanged.  In particular  this group does not, in general preserve
the space of potential supercurrents:
\be
\CV_{1^6} \otimes \tilde \CV_{0^6} \oplus  \CV_{0^6} \otimes \tilde \CV_{1^6}
\ee
The subgroup
that preserves  $\CH_{GTVW}^{3/2,0} \oplus \CH_{GTVW}^{(0,3/2)} $ is that
where  image of the determinant map is either all $+1$ or all $-1$.
We will denote this group as:
\be
{\rm P}\left( {\rm in}(4) \right)^6
\ee
Put simply: The spinor lift of a parity transformation is diagonally embedded in all
six factors. Thus the group ${\rm P}\left( {\rm in}(4) \right)^6 : S_6$
acts both on the space $\CH_{GTVW}^{3/2,0} \oplus \CH_{GTVW}^{(0,3/2)} $
as well as on the RR ground states.

\subsubsection{Stabilizer Of $(4,4)$ Supersymmetry Within ${\rm Spin}(4)^6 : S_6/Z_{\rm diag}$}\label{subsec:44Preserving}

We are now in a good position to determine the group of symmetries of the GTVW model
that commute with $(4,4)$ supersymmetry. One way to construct such symmetries proceeds
by lifting suitable subgroups of the holomorph of the hexacode to $\left({\rm Spin}(4)^6/Z_{\rm diag}\right): S_6$.
We will construct a group of symmetries  that is related to $M_{20}$ in a way
explained in detail below. Since the discussion has several subtleties we will be
going into excruciating and explicit detail.

To begin, we work chirally, and consider lifts of holomorphs of the hexacode to the
semidirect product
 $SU(2)^6 : S_6$ which preserve all four chiral supersymmetries.
 Now, elements of $SU(2)^6 : S_6$ that commute with all four chiral supersymmetries
 must  commute with the $SU(2)$ R-symmetry. In the GTVW model the chiral R-symmetry
 is identified with  the first $SU(2)$ factor in the product $SU(2)^6$. Symmetries of this type
that are lifts of holomorphs of the hexacode must be lifts of holomorphs
that preserve the first digit of the hexacode. Therefore we begin by determining the subgroup, $F\subset \Aut(\CH_6)$,
that fixes the first digit of the hexacode.
%
%
%
It is easy to see that $F\subset H_5 = \langle g_1, \dots, g_5 \rangle$.
 The elements $g_2=(34)(56)$ and $g_4=(35)(46)$
of  Appendix \ref{app:HexacodeAutGroup} are certainly in $F$
As an example of a nontrivial element of $F$ we can modify the generator $g_5$ in
 Appendix \ref{app:HexacodeAutGroup} by combining with elements of $H_0 \times H_3$ to get, for example:
\be
g_5' :  (x_1, \dots, x_6) \mapsto (x_1, x_5 , \uw x_4, \uwb x_3, x_2, x_6)
\ee
Now recall (equation \eqref{eq:TrackPerm}) that there is a homomorphism $p:\Aut(\CH_6) \to S_6$
that simply tracks what permutation of digits the automorphism implements. We have
$p(g_5') =  (25)(34)$. Now $(34)(56)$ and $(35)(46)$ generate a $\IZ_2 \times \IZ_2$ subgroup of order 4.
But $(34)(56)(25)(34)=(256)$ has order $3$ while  $(35)(46) (25)(34)=(23645)$ has order $5$. It follows that
$4 \cdot 3 \cdot 5 = 60$ must divide the order of $p(F)$. On the other hand, $p(F)$ is a proper subgroup of $S_5$
and hence $p(F)$ must be exactly $A_5\subset S_5$. The kernel of $p$ restricted to $F$ is trivial so
$F$ is isomorphic to $A_5$.

Now, still working chirally,  let us consider the ``translation symmetries,'' that is, the
action by $SU(2)_6$ elements $h(w)$ (acting either on the left- or the right- movers)
that commute with $N=4$ supersymmetry. Thanks to the description of the $N=4$ currents in
section \ref{subsec:N=4AndCodes} we see that these translations by $w \in \CH_6$ must commute
with $P_x$ and hence must preserve the first digit of hexacode words. Therefore when
acting with $h(w)$ the first digit of $w$ must be $\uz$.
Let   $\CH_6^0\subset \CH_6$ be the subspace consisting of hexacode words whose first digit is $\uz$.
 As we have seen, this is a $2$-dimensional subspace
over $\IF_4$. It has $16$ elements and, as an Abelian group,   $\CH_6^0 \cong \IZ_2^4$.
Since we can independently lift a translation by a hexacode word to
elements of both $SU(2)_L^6$ and $SU(2)_R^6$ we are led to consider the group
\be\label{eq:M20-appears}
 (\CH_{6,L}^0 \times \CH_{6,R}^0):F
\ee
with $F$ acting diagonally as a group of automorphisms.

Now we must lift these group operations to $\left({\rm Spin}(4)^6/Z_{\rm diag}\right): S_6$.
First,  as we have seen, the lift $\langle \widehat g_1, \dots, \widehat{g}'_5 \rangle$
does not generate a central extension in $SU(2)_6:S_6$ and
we will demonstrate below that the further lift to $SU(2)^6:S_6$ is isomorphic to $A_5$.

At this point, the reader should recall the important remark concerning the relation of
$SU(2)^6$ to $SU(2)_6$ mentioned in section \ref{subsec:HexaRepSuperCurrent}.

In order to lift to $\left({\rm Spin}(4)^6/Z_{\rm diag}\right): S_6$ note that
if $p(\varphi)$ is a nontrivial element of $S_6$ then it must act {\it diagonally} on
the left- and right-movers.
If $\varphi \in F$ we will denote its lift to the full GTVW model by $\widehat\varphi$.
As discussed near \eqref{eq:LiftCH6}, when lifting $\CH_6$  to
a subgroup of $SU(2)^6$ via
\be
(x_1, \dots, x_6) \rightarrow (h(x_1), \dots, h(x_6)) \in SU(2)^6
\ee
 we encounter an extension by $Z \cong \IZ_2^5$. Again, we stress that
 the cocycle defined by this section is a $6$-tuple of the cocycles
 \eqref{eq:h-cocycle-def}, and not their product!  Altogether then,
 we have a group $\CG_{4,4}$ which fits in an extension:
\be\label{eq:G44-structure}
1\rightarrow  Z \rightarrow \CG_{4,4} { \buildrel \pi \over \longrightarrow}  \left( \CH_{6,L}^0 \times \CH_{6,R}^0\right):F  \rightarrow 1
\ee
where $Z$ should really be regarded as $\left( Z_L \times Z_R \right)/Z_{\rm diag}$.  More abstractly, $\CG_{4,4}$
has the structure of a nontrivial central extension:
\be
\CG_{4,4} \cong \IZ_2^5 \cdot \left(  \left( \IZ_2^4 \times \IZ_2^4 \right): A_5 \right)~ .
\ee
Note that
\be\label{eq:G44-order}
\vert \CG_{4,4} \vert = 2^{15}\cdot 3 \cdot 5
\ee
We will now argue that the central extension \eqref{eq:G44-structure} is nontrivial.

It is possible to give a very concrete description of the group $\CG_{4,4}$.
We will write group elements in $\left( SU(2)^6_L \times SU(2)^6_R) \right):S_6$ using the notation
\be
\left( \left(u_1, \dots, u_6 \right)_L , \left( \tilde u_1, \dots, \tilde u_6\right)_R ; \sigma\right)
\ee
with $u_i, \tilde u_i \in SU(2)$ and $\sigma \in S_6$. The multiplication rule is the usual
semidirect product rule. As mentioned above, to obtain
a group acting effectively on the CFT we must take a quotient
by $Z_{\rm diag}$. In other words it is understood that we identify:
\be
\left( \left(u_1, \dots, u_6 \right)_L , \left( \tilde u_1, \dots, \tilde u_6\right)_R ; \sigma\right)
\sim
\left( \left(z_1 u_1, \dots,z_6 u_6 \right)_L , \left( z_1 \tilde u_1, \dots, z_6 \tilde u_6\right)_R ; \sigma\right)
\ee
where $(z_1, \dots, z_6 ) \in Z \subset Z(SU(2)^6)$.

Now we choose a section of $\pi$ over  $F$ in equation \eqref{eq:G44-structure}  generated by
\be
\begin{split}
\widehat{g}_2 & = \left( 1^6_L , 1^6_R ; (34)(56) \right)\\
\widehat{g}_4 & = \left( 1^6_L , 1^6_R ; (35)(46) \right)\\
\widehat{g}'_5 & =
\left( \left(  \left( 1, 1, \Omega^{-1}, \Omega,1,1 \right)_L , \left( 1, 1, \Omega^{-1}, \Omega,1,1 \right)_R \right); (25)(34) \right) \\
\end{split}
\ee
This section splits the sequence over $F$ and defines a subgroup of $\CG_{4,4}$ isomorphic to $A_5$.
 It is an amusing exercise to verify that $\widehat{g}'_5$ has order two, that $\widehat{g}_2\widehat{g}'_5$ has order $3$,  and
that $\widehat{g}_4\widehat{g}'_5$ has order $5$.

Now, over $\CH_{6,L} \times \CH_{6,R}$ we choose the section
\be
\left( \left(h(x_1), \dots, h(x_6) \right)_L , \left( h(\tilde x_1), \dots, h(\tilde x_6) \right)_R ; 1\right)
\ee
where $(x_1, \dots, x_6) \in \CH_6$ and $(\tilde x_1, \dots, \tilde x_6) \in \CH_6$. The restriction
to $\CH_{6,L}^0 \times \CH_{6,R}^0$ imposes $x_1 = \tilde x_1 = \uz$ and consequently $h(x_1) = h(\tilde x_1) = 1$.
The multiplication of these group elements will clearly involve a $6$-tuple of cocycles $\epsilon(x,y)$ in
equation \eqref{eq:h-cocycle-def} and hence the extension by $Z$ might be nontrivial.

There are two
ways to see that the extension \eqref{eq:G44-structure} is indeed nontrivial. First, note that every element of
$\CH_{6,L}^0 \times \CH_{6,R}^0$ is an involution. Consider the square of the lift of any nonidentity
element in $\CH_{6,L}^0 \times \CH_{6,R}^0$. The square will be a nonidentity element in the  subgroup $Z^0 \subset Z$
defined by the condition $z_1 = 1$. In fact, all elements of $Z^0$ can be obtained in this way.
(Note that $Z^0$ is isomorphic to $\IZ_2^4$.)
On the other hand, every element in $Z$ is order two so it is impossible to produce a nonidentity
element which is a perfect square. Therefore there exist involutions whose lifts
have squares which are not themselves perfect squares in $Z$. This implies the extension is nontrivial.
\footnote{For an explanation of this result the reader might wish to consult Remark 5 in section 14.3 of \cite{MooreCourse}. }
In fact, every nonidentity element in $\CH_{6,L}^0 \times \CH_{6,R}^0$ provides an example.

A second way to see that the extension \eqref{eq:G44-structure} is nontrivial is to note that since
$\CH_{6,L}^0 \times \CH_{6,R}^0$  is Abelian it suffices to check if the commutator of
lifted group elements is trivial or not.   The story is the same for left- and right-movers so we might as well take
$\left( h(\tilde x_1), \dots, h(\tilde x_6) \right)_R = 1^6_R$. Then the commutator of the
lifts of elements in $\CH_{6,L}^0$ is of the form:
\be
\left( \left( c(x_1, y_1), \dots, c(x_6,y_6) \right)_L \times 1^6_R ; 1 \right)
\ee
where $(x_1, \dots,x_6) \in \CH_6^0$ and  $(y_1, \dots,y_6) \in \CH_6^0$ and the
commutator function is
\be
c(x,y) = \begin{cases}  1 & x=y \quad {\rm or} \quad xy =\uz \\
 -1  & x \not= y \quad {\rm and} \quad xy \not= \uz \\
\end{cases}
\ee
Again, a simple perusal of equation \eqref{eqn:Hexacode} shows that every element of $Z^0$ appears as a group commutator.

We can now describe the relation to the group $2^8:M_{20}$ obtained in \cite{gtvw}.
Although the extension \eqref{eq:G44-structure} is a nontrivial central extension
it is in fact isomorphic to $2^9:M_{20}$ where the $2^9$ is a \underline{noncentral} subgroup.
\footnote{The relevant isomorphism was discovered by T.~ Johnson-Freyd, and we
thank him for extensive discussions and clarifications related to this.}
For each $v\in \IF_4^*$ we will define sub\underline{groups} $\CH_{LR}(v)$ of $(Q^6 \times Q^6)/Z_{\rm diag}$.
Each subgroup $\CH_{LR}(v)$ is isomorphic to  $\CH_6$. The group $\CH_{LR}(v)$ is the group of elements
\be\label{eq:SkewEmbed}
[(h(x_1,\dots, h(x_6))_L, (h(v x_1), \dots, h(v x_6))_R ]
\ee
where the square brackets denote the equivalence class under the quotient by $Z_{\rm diag}$.
Note that the group elements $(h(x_1), \dots, h(x_6) ) \in Q^6$ do \underline{not} form a
subgroup, even when we restrict to $(x_1, \dots, x_6 ) \in \CH_6$. However the group
elements \eqref{eq:SkewEmbed} do form a subgroup thanks to the division by $Z_{\rm diag}$ and
the permutation invariance of the cocycle (see equation  \eqref{eq:CocycPermInv}). Again, using permutation
invariance of the cocycle, the elements of the form \eqref{eq:SkewEmbed} are all
involutions. Indeed, $\CH_{LR}(v) \cong \IZ_2^6$ as an Abelian group.
Similarly, the restriction to the
subgroup with $x_1=0$ defines subgroups $\CH_{LR}^0(v)$ of $\CG_{4,4}$, each of which is
isomorphic to $\IZ_2^4$. Now it is easy to check that  $\CH_{LR}(v)$ acts, via conjugation,  on
$\CH_{LR}(v')$ for $v\not= v'$ as a nontrivial automorphism, and similarly for
$\CH_{LR}^0(v)$ and $\CH_{LR}^0(v')$ for $v\not= v'$.
Now denote by $\CG_{4,4}^1\subset \CG_{4,4}$ the subgroup defined by the fiber of $\pi$ over the
``part with $F=1$''.
\footnote{To be more precise: Let $\pi^{(2)}:= p^{(2)}\circ \pi$ where
 $p^{(2)}: \left( \CH_{6,L}^0 \times \CH_{6,R}^0\right):F \to F$ is the projection
 and define $\CG_{4,4}^1$ to be the kernel of $\pi^{(2)}$. }
The center of $ \CG_{4,4}^1$ is $Z$. Each of the subgroups $ Z \times \CH_{LR}^0(v') \subset \CG_{4,4}^1$
is a normal subgroup. A complementary group can be taken to be $\CH_{LR}(v)$ with $v' \not=v$, that is
we can write
\be
\CG_{4,4}^1 \cong \left( Z \times \CH_{LR}^0(v') \right): \CH_{LR}(v)
\ee
for any pair $v\not=v'$, where in the semidirect product $\CH_{LR}(v)$ acts nontrivially on $\CH_{LR}(v')$ by conjugation.
\footnote{To make this completely explicit, the main point is to note that for all $a$ and $b$ we can solve
\be
\begin{split}
[\left( h(a_1), \dots, h(a_6) \right)_L , \left( h(b_1), \dots, h(b_6) \right)_R ]
& = \\
z\cdot [\left( h(x_1), \dots, h(x_6) \right)_L , \left( h(v x_1), \dots, h(v x_6) \right)_R ]
&
\cdot [\left( h(y_1), \dots, h(y_6) \right)_L , \left( h(v' y_1), \dots, h(v' y_6) \right)_R ] \\
\end{split}
\ee
for some $z\in Z$ and $x,y\in \CH_6$. To see this note that
 for each $\alpha$ we need to solve $x_\alpha + y_\alpha = a_\alpha$ and
$v x_\alpha + v'  y_\alpha = b_\alpha$. One can easily check that for $v,v' \in \IF_4^*$
with $v\not=v'$ there exists a solution.
}

%
%
%
%
Now, since the fibration trivializes over $F$ we can restore $F$ to write
\be\label{eq:Semidirect-vvprime}
\CG_{4,4} \cong \left( Z \times \CH_{LR}^0(v') \right): \left( \CH_{LR}(v):F \right)
\ee
We can now make contact with the group $M_{20}$. It is known
\footnote{See http://brauer.maths.qmul.ac.uk/Atlas/v3/misc/M20/ .}
that
\be
M_{20} \cong \IZ_2^4: A_5
\ee
and so we can identify
\be\label{eq:M20-appears-prime}
M_{20} \cong \CH_6^0(v):F ~.
\ee
for any $v\in \IF_4^*$. On the other hand,
\be
Z \times \CH_6^0(v') \cong \IZ_2^9
\ee
and hence we obtain
\be
\CG_{4,4} \cong \IZ_2^9:M_{20}
\ee
where only $\IZ_2^5 \subset \IZ_2^9$ is central.

Finally, we comment on the difference of $2^9:M_{20}$ vs. the group $2^8:M_{20}$ that
appears in \cite{gtvw}. The space of RR groundstates decomposes, under the $SU(2)_L \times SU(2)_R$
$R$-symmetry group as $(\textbf{2}; \textbf{2}) \oplus 20 (\textbf{1}; \textbf{1})$
and the theorem of \cite{Gaberdiel:2011fg} only addresses the commutant of $(4,4)$
supersymmetry that acts trivially on the subspace $(\textbf{2}; \textbf{2}) $ of RR states.
Comparing with equation \eqref{eq:gtvw-RR-gndstates} for the RR groundstates in the
GTVW model we see that to compare results we should only consider the subgroup of $\CG_{4,4}$
that acts trivially on the space $(\textbf{2}; \textbf{2})_{\alpha=1}$. This subgroup,
$\CG_{4,4}^0$ is obtained by restriction of the semidirect product
\eqref{eq:Semidirect-vvprime} obtained by replacing $Z$ with the subgroup  $Z^0\subset Z$
with $z_1=1$. That subgroup is isomorphic to $\IZ_2^4$.  The ``extra'' $\IZ_2$
can be generated by $\left((-1^6_L, 1^6_R);1\right)$ and this element can be interpreted as
$(-1)^{F_R}$. Thus,
\be
\CG_{4,4}^0 \cong 2^8:M_{20}
\ee
in accord with \cite{gtvw}. Note again that the $2^8$ is noncentral.

\bigskip
\noindent
\textbf{Remarks}

\begin{enumerate}

\item Here is a slightly more conceptual description of the
group $\CG_{4,4}$. It is included for the benefit of fussbudgets.
The subgroup of the holomorph of the hexacode that preserves the first
hexacode digit is the semidirect product $\CH_6^0 : F $. Its lift
to $SU(2)^6 : S_6$ defines a subgroup $\widetilde{K}$ that
fits in an exact sequence
\be
1 \rightarrow Z \rightarrow \widetilde{K} \rightarrow \CH_6^0 : F \rightarrow 0
\ee
because, as we have seen,  the lift of elements in $\langle g_0, \dots, g_5 \rangle$
act without central extension.
When combining   left-movers with right-movers we aim to produce a
subgroup of ${\rm Spin}(4)^6 : S_6$ where group elements that involve nontrivial
permutations of factors must act diagonally on left- and right-movers.
We therefore view ${\rm Spin}(4)^6 : S_6$ as a fiber product
of $\left( SU(2)^6 : S_6\right)_L \times_{p_L, p_R}  \left(SU(2)^6 : S_6\right)_R$
where $p$ is the projection to $S_6$.
\footnote{Recall that given groups and homomorphisms $\psi_1: G_1 \to H$ and
$\psi_2: G_2 \to H$ the fiber product is
\be
G_1 \times_{\psi_1, \psi_2} G_2 = \{ (g_1, g_2) \vert \psi_1(g_1) = \psi_2(g_2) \} ~  .
\ee
}
Our symmetry group will be a quotient of the fiber product
$\widetilde{K}_L \times_{p_L, p_R } \widetilde{K}_R$. The
reason we must take a quotient is that  $Z_L \times Z_R$ does not
act effectively on the GTVW spectrum. The reason is that
\be
(\epsilon_1, \dots, \epsilon_6) \subset Z
\ee
acts on $\CV_s$ via the scalar $\prod_{\alpha}  \epsilon_{\alpha}^{s_{\alpha}} $.
But then the diagonally embedded subgroup   $Z_{\rm diag} \subset Z_L \times Z_R$ acts
on $\CV_s \otimes \widetilde{\CV}_{\tilde s}$ as
\be
\prod_{\alpha=1}^6  \epsilon_{\alpha}^{s_\alpha + \tilde s_{\alpha} }
\ee
If $\tilde s_{\alpha} = s_{\alpha} $ this factor is equal to $1$. If $\tilde s_{\alpha} = s_{\alpha} +1$
the factor is $\prod_{\alpha} \epsilon_\alpha = 1$, by the definition of $Z$. One can check that $Z_{\rm diag}$
is the largest subgroup of $Z_L \times Z_R$ that acts ineffectively. Thus, the group of symmetries
preserving $(4,4)$ supersymmetry that we have identified is properly described as
\be\label{eq:G44-Canon}
\CG_{4,4} \cong \left( \widetilde{K}_L \times_{p_L, p_R}  \widetilde{K}_R \right)/Z_{\rm diag}
\ee

\item It is also instructive to compare our description of the generators of the
group of $(4,4)$ preserving symmetries with the specific transformations
studied in   \cite{gtvw}. The $S_6$ elements  $(34)(56)$ and $(35)(46)$ above correspond to the transformations
$s_{v_4}$ and $s_u$ respectively in \cite{gtvw} and have a geometrical origin as half period shifts in the
$T_{D_4}/{\mathbb Z}_2$ orbifold description of the GTVW sigma model. See their equation $(4.34)$. The $S_6$ permutation
$(25)(34)$ corresponds to $\alpha^{p,T}$ and while less obvious is also a half-period shift. See their equation $(4.62)$.
This lifts to the element $\widehat{g}'_5$
which is still an involution. Thus these elements generate a group isomorphic to $A_5$.
The symmetries in equation $(4.41)$ of \cite{gtvw} correspond to (the lift of) $\CH_{6,L}^0$.
Since they are purely left-moving they are
nongeometric symmetries.  See the discussion at the end of sec. 4.2 of \cite{gtvw}.
Our group $Z$ corresponds to the symmetries denoted $t_i t_j$ in \cite{gtvw}.
The symmetries  $\gamma_1$ and $\gamma_2$, corresponding to  rotations acting on the Kummer surface
$T_{D_4}/{\mathbb Z}_2$ and the symmetries $s_{v_1+v_2}$, $s_{v_2+v_4}$ corresponding to
or half period shifts in the $D_4$ lattice are related in equation $(4.39)$ of \cite{gtvw} to
diagonally embedded elements of $\CH_6^0$.  Our description is evidently less geometric, but has the
benefit of unifying the treatment of symmetries in terms of the holomorph of the hexacode.

\end{enumerate}

\subsubsection{Stabilizer Of $(4,1)$ Supersymmetry Within ${\rm Spin}(4)^6 : S_6/Z_{\rm diag}$}\label{subsec:41Preserving}

The ideas of section \ref{subsec:44Preserving} can readily be generalized to produce a group of
symmetries of the GTVW model that commute with $(4,1)$ supersymmetry. As discussed above, these
should still be relevant to the degeneracies computed by the elliptic genus.

If we only aim to preserve a single $N=1$ right-moving supercurrent then we can drop the
restriction that our symmetries act trivially on the first hexacode digit.
Using the canonical description of \eqref{eq:G44-Canon} we replace $\widetilde{K}_R$ by the lift
of the holomorph of the hexacode to   $SU(2)^6:S_6$.
Specifically we now include elements $\varphi \in \langle g_0 \rangle \times F \subset \Aut(\CH_6)$
as well as arbitrary translations $w_R\in \CH_6$. Call the lift $\widetilde{K}'_R$. The analog of
\eqref{eq:G44-Canon} is then
\be\label{eq:G41-Canon}
\CG_{4,1} = \left( \widetilde{K}_L \times_{p_L, p_R}  \widetilde{K}'_R \right)/Z_{\rm diag}
\ee
The group will have structure analogous to \eqref{eq:G44-structure}:
\be\label{eq:G41-structure}
\CG_{4,1} \cong Z \cdot \left( \left(  \CH_{6,L}^0 \times \CH_{6,R}\right):F \right)
\ee
Note that we cannot make use of other
automorphisms $\varphi_R$ in $\langle g_1, \dots, g_5, g_F \rangle -F$ because these have permutations that
change the first digit. Because of the fiber product structure (saying that the  permutation image of
the action of $\varphi$ on the left-movers and right-movers must be the same) such automorphisms
of the CFT do not commute with the SU(2) R-symmetry of the {\it left}-moving N=4 superconformal
algebra.

Note that  $\CG_{4,4}$ is a normal subgroup of $\CG_{4,1}$ and
\be
\CG_{4,1}/\CG_{4,4} \cong \left( \CH_{6,R}/\CH_{6,R}^0 \right): \langle \widehat{g_{0}}_R \rangle/\langle -1^6 \rangle
\cong \left( \IZ_2 \times \IZ_2\right): \IZ_3
\ee
In this sense the group $\CG_{4,1}$ is $12$ times bigger. In particular,
\be
\vert \CG_{4,1} \vert = 2^{17}\cdot 3^2 \cdot 5
\ee
For comparison note that
\be
\vert M_{24} \vert = 2^{10}\cdot 3^3 \cdot 5 \cdot 7 \cdot 11 \cdot 23
\ee
So,  $\CG_{4,1}$ cannot have $M_{24}$ as a quotient group, by Lagrange's theorem.

\subsection{The ``New'' Twined Elliptic Genera}\label{subsec:NewTwined}

In this section we define a few elliptic genera associated to some
of the ``new'' elements in $\CG_{4,1}$. We will not give a systematic
study of the full vector space of such ``new'' elliptic genera.

If $g$ is an automorphism in $\CG_{4,1}$ we can define a twisted elliptic genus:
\be\label{eq:TrueTwistedEG}
\CE_g(z,\tau):= \Tr_{\CH_{RR}} (-1)^{F_L + F_R}U(g)  e^{2\pi \I z J_0} q^{H} \bar q^{\tilde H}
\ee
where $J_0 = 2J_{0}^3$ is normalized to have integral eigenvalues and $U(g)$ is the action on
the GTVW space of states.  The supersymmetric
cancellations will continue to hold so we get a Jacobi form of weight zero and index 1 for a
suitable congruence subgroup.

In order to compute \eqref{eq:TrueTwistedEG} we begin by isolating the subspace of $\CH_{RR}$ that contains R-moving groundstates,
that is, the subspace of the GTVW state spacethat contains RR sector states with
$(h,\tilde h) = (h, \frac{1}{4})$.  To this end, for $1\leq \alpha \leq 6$ define
\be
\CV^{(\alpha)} = \CV_{s_1} \otimes \cdots \CV_{s_6}
\ee
where $s_{\alpha}=1$ and $s_{\beta}=0$ for $\beta \not= \alpha$. Similarly, define
\be
\CV^{(\alpha)+e} = \CV_{s_1} \otimes \cdots \CV_{s_6}
\ee
where $s_{\alpha}=0$ and $s_{\beta}=1$ for $\beta \not= \alpha$.
The relevant subspace of the RR states is then
\be\label{eq:Relevant}
\oplus_{\alpha=1}^6 \left[ \CV^{(\alpha)} \otimes \widetilde{\CV}^{(\alpha)}
\oplus \CV^{(\alpha)+e} \otimes \widetilde{\CV}^{(\alpha)}\right]
\ee
Note that $(-1)^{F_L + F_R}$ is $+1$ on the first summand and $-1$ on the
second summand in the expression in square brackets above.

Now, for a single Gaussian model we have
\be
\Tr_{\CV_0} e^{2\pi \I z J_0} q^{H} = \frac{ \vartheta_3(2z,2\tau)}{\eta(\tau)} := f_0(z)
\ee
\be
\Tr_{\CV_1} e^{2\pi \I z J_0} q^{H} = \frac{ \vartheta_2(2z,2\tau)}{\eta(\tau)} := f_1(z)
\ee
also let $f_0:=f_0(0)$ and $f_1 := f_1(0)$.
The twisted elliptic genera for group elements with $g^L=1$
are in the ring of functions of $(z,\tau)$ generated by $f_0(z), f_1(z), f_0,f_1$.
In fact, they will be in the linear span of the functions:
\be
\begin{split}
F_0(z,\tau) & :=  f_1(z) f_0^5 -  f_0(z) f_1^5 \\
F_1(z,\tau) & :=  f_0(z) f_1 f_0^4 - f_1(z) f_0 f_1^4 \\
\end{split}
\ee
For example, the elliptic genus itself is just
\be
\CE_1(z,\tau) = 2 (F_0 + 5 F_1)
\ee

Now, for $x\in \IF_4$ we choose coset representatives of $\CH_6/\CH_6^0$. We will
make the explicit choice:
\be
\begin{split}
w_{\uz} & := (\uz,\uz,\uz,\uz,\uz,\uz)\\
w_{\uone} & := (\uone,\uone,\uw,\uw,\uwb,\uwb) \\
w_{\uw} & := (\uw,\uw,\uwb,\uwb,\uone,\uone) \\
w_{\uwb} & := (\uwb,\uwb,\uone,\uone,\uw,\uw) \\
\end{split}
\ee

For $a\in \IZ_6$ and $x\in \IF_4$ define:
\be
\CE_g^{a,x}(z,\tau) :=  \CE_{g\cdot ( 1; \hat g_0^{-a} h(w_x) )}(z,\tau)
\ee
Note that $\CE_g^{a+3,x}(z,\tau) = - \CE_g^{a,x}(z,\tau) $. If we apply this
formula to $g\in \CG_{4,4}$ then we define, in principle, $12-1 = 11$ ``new''
elliptic genera for each of the ``old'' elliptic genera.

A small computation leads to the following table of ``new'' twisted elliptic genera
for the case $g=1$:
\vskip1in
\begin{center}
\begin{tabular}{|c|c|c|c|c|}
\hline
    & $x= \uz$ & $x=\uone$ & $x=\uw$ & $x=\uwb$    \\
\hline $a=0 $&  $2(F_0 + F_1) $ & $0$ & $4 F_1 $ & $4 F_1 $   \\
\hline  $a=1$ & $F_0 - 3 F_1 $& $-(F_0 + 5 F_1) $  &  $ - F_0 - F_1   $ & $ - F_0  - F_1   $ \\
\hline  $a=2$ & $ -(F_0 + 5 F_1)  $& $-(F_0 + 5 F_1) $  &  $-(F_0 + 5 F_1)  $ & $-(F_0 + 5 F_1)  $ \\
\hline \end{tabular}\end{center}
\vskip1in

One might wonder whether the new group elements we have found are
related to a subquotient of the still-mysterious $M_{24}$ symmetry
of K3 sigma models. We will argue now that they are not.
Recall the discussion near equation \eqref{eq:H-type-reps}.
The representations of $M_{24}$,  $H_{0,0}, H_{0,1/2}, H_n $, with $n\geq 1$
have the property that, for {\it every} $g\in M_{24}$,
the function $\hat \phi_g(z,\tau)$ defined in \eqref{eq:PseudoEG}
%
%
transform as Jacobi forms - precisely analogous to those in \eqref{eq:TrueTwistedEG},  {\it as if} $g \in M_{24}$ acted on
the CFT $\CC$  as a $(4,4)$-preserving automorphism. Moreover as noted in \eqref{eq:Equality}, for $g\in \CG_{4,4}$ they indeed coincide.
%
%

The first few representations $H_{0,0}, H_{0,1/2}$ and $H_n \qquad n\geq 1$
determined by \cite{Eguchi:2010ej, Cheng:2010pq, Gaberdiel:2010ca, Eguchi:2010fg, Gannon} are
\be
\begin{split}
H_{0,0} & = \textbf{23} - 3 \cdot \textbf{1} \\
H_{0,1/2} & = - 2\cdot \textbf{1} \\
H_1 & =  \textbf{45} \oplus \textbf{45}^* \\
H_2 & =  \textbf{231} \oplus \textbf{231}^* \\
H_3 & =  \textbf{770} \oplus \textbf{770}^* \\
\end{split}
\ee

We now explain that the twined genera associated to the ``new'' elements in $\CG_{4,1} - \CG_{4,4}$
do {\it not} conform to the expectation   \eqref{eq:Equality} by looking at a few examples.
We begin by noting the decomposition of $F_0$ and $F_1$ into $N=4$ characters:
\be\label{eq:F0-cd}
F_0 = -\chi_{0,1/2}-15 \chi_{5/4}-49 \chi_{9/4}-210 \chi_{13/4}-543 \chi_{17/4}-1484 \chi_{21/4} + \cdots
\ee
\be\label{eq:F1-cd}
F_1 = 2 \chi_{0,0}+12 \chi_{5/4}+56 \chi_{9/4}+196 \chi_{13/4}+564 \chi_{17/4}+1456 \chi_{21/4}
\ee

\begin{enumerate}

\item Now note that for $g=(1; (\Omega^{\otimes 6})^a h(w_x + (001111)))$  we find
$- (F_0 + 5 F_1)$. If our ``new'' $g$ were an $M_{24}$ element then we would
need to have $\Tr_{H_n}(g) = - \half \dim H_n$. There is no such group element.

\item Next consider $a=0, x= \uw$ and $a=0, x=\uwb$. These give $4F_1$. Looking at \eqref{eq:F1-cd} we see
the coefficient of $ch_{0,1/2}$ is zero. On the other hand,  $H_{0,1/2} = -2 \textbf{1}$
and the character of every element $g$ in this representation is $-2$. This shows
we have not made some error by confusing traces with supertraces.

\item In some discussions of Moonshine authors will distinguish massless and massive states.
Let us consider again the character   $4 F_1$. We would need to find a $g \in M_{24}$
with
\be
\Tr_{\textbf{45} \oplus \textbf{45}^*}(g) = 4\cdot 12
\ee
\be
\Tr_{\textbf{231} \oplus \textbf{231}^*}(g) = 4\cdot 56
\ee
it is easy to see from the character table of $M_{24}$  that no such $g$ exists.

\item Similarly, for $F_0 - 3 F_1$ we would require, at the massive level
\be
\Tr_{\textbf{45} \oplus \textbf{45}^*}(g) = -51
\ee
\be
\Tr_{\textbf{231} \oplus \textbf{231}^*}(g) = -217
\ee
Again, no such $g$ exists.

\end{enumerate}

\subsection{Stabilizer Of An $N=1$ Supercurrent Within A Group That Includes Parity-Reversing Operations }

As we will see in the next section, the full structure of the Golay code as a
symmetry of the RR groundstates of the GTVW model only becomes apparent when
we combine left- and right-moving supersymmetries and study the stabilizer,
within ${\rm Pin}(4)^6:S_6$ of the $N=1$ supercurrent based on
\be
\Psi_L - \Psi_R
\ee
The potential supercharges live in the subspace of the GTVW space of states:
\be
\CH_{GTVW}^{3/2,0} \oplus \CH_{GTVW}^{(0,3/2)} \cong
\otimes_{\alpha=1}^6 \left( \textbf{2} \right)_L^{(\alpha)}   \oplus \otimes_{\alpha=1}^6 \left( \textbf{2}  \right)_R^{(\alpha)}
\ee
In order to see the full symmetry of the RR states we need to extend the quaternion group $Q^6$ used in
section \ref{subsubsec:StabInQ6} to include parity. We do this by extending
$SU(2)_L \times SU(2)_R$ to ${\rm Pin(4)}$ and viewing ${\rm Pin}(4) \subset \IH(2)$, the
algebra of $2\times 2$ matrices over the quaternions.  Viewed this way we are lead to consider
a group consisting of elements
\be
\left(  \begin{pmatrix} \epsilon_L^1  h(x_1) & 0  \\  0 &  \epsilon_R^1  h(x_1) \\ \end{pmatrix}, \dots,
\begin{pmatrix} \epsilon_L^6  h(x_6) & 0  \\  0 &  \epsilon_R^6  h(x_6) \\ \end{pmatrix} \right)
\ee
together with
\be
\left(  \begin{pmatrix} 0 & \epsilon_L^1  h(x_1)   \\   \epsilon_R^1  h(x_1) & 0 \\ \end{pmatrix}, \dots,
\begin{pmatrix}0 &  \epsilon_L^6  h(x_6)  \\    \epsilon_R^6  h(x_6) & 0  \\ \end{pmatrix}\right)
\ee
where $x_i \in \IF_4$.
This is a non-abelian group of order $2^{25}$ and we denote it by
$\widehat{Q}^6_{P}$. The subscript $P$ indicates that we have included a diagonally-acting
parity operation.  The above group acts   naturally on $\CH_{GTVW}^{3/2,0} \oplus \CH_{GTVW}^{0,3/2}$.
The stabilizer of the $N=1$ supercurrent determined by $\Psi_L - \Psi_R$  within $\widehat{Q}^6_{P}$
can be determined to be:
\be\label{eq:StabQhatP}
\begin{split}
{\rm Stab}_{\widehat{Q}^6_{P} } (\Psi_L - \Psi_R)&  = \{
\left(  \begin{pmatrix} \epsilon_L^1  h(x_1) & 0  \\  0 &  \epsilon_R^1  h(x_1) \\ \end{pmatrix}, \dots,
\begin{pmatrix} \epsilon_L^6  h(x_6) & 0  \\  0 &  \epsilon_R^6  h(x_6) \\ \end{pmatrix} \right)  \\
&
 \vert \prod_{i=1}^6 \epsilon_L^i = 1 \quad \& \quad  \prod_{i=1}^6 \epsilon_R^i = 1 \quad \& \quad   (x_1, \dots, x_6 ) \in \CH_6 \}\\
 & \amalg \\
 &
  \{ \left(  \begin{pmatrix} 0 & \epsilon_L^1  h(x_1)   \\   \epsilon_R^1  h(x_1) & 0 \\ \end{pmatrix}, \dots,
\begin{pmatrix}0 &  \epsilon_L^6  h(x_6)  \\    \epsilon_R^6  h(x_6) & 0  \\ \end{pmatrix}\right) \\
&
 \vert \prod_{i=1}^6 \epsilon_L^i =- 1 \quad \& \quad  \prod_{i=1}^6 \epsilon_R^i = - 1 \quad \& \quad   (x_1, \dots, x_6 ) \in \CH_6 \}\\
\end{split}
\ee
This is a non-abelian group of order $2^5 \times 2^5 \times 4^3 \times 2$. The main significance of this group will be apparent when
we consider its action on the RR ground states.

We should note that above we have only discussed the ``translation symmetries'' of $\Psi_L - \Psi_R$.
There will also be a larger group making use of the  ``rotational'' automorphisms of the hexacode.
We have not explored this larger symmetry in detail.

\section{RR States And The MOG Construction Of The Golay Code} \label{sec:MOG}

In this section we will show that the group \eqref{eq:StabQhatP} that preserves
the $N=1$ supercurrent $\Psi_L - \Psi_R$ also acts on   the space of RR ground states according
to a pattern governed by the Golay code. The pattern emerges when we use a special
basis of RR states, so we be begin by explaining this distinguished basis.

The space of RR ground states,
as an $(SU(2)_L \times SU(2)_R)^6$ representation,
has the structure:
\be\label{eq:gtvw-RR-gndstates}
\CV_{RR} = \CH_{GTVW}^{1/4,1/4} \cong \oplus_{\alpha=1}^6 \left( \textbf{2} ; \textbf{2}  \right)^{(\alpha)}
\ee
Now the representation $\left( \textbf{2} ;\textbf{2}  \right)$ of $SU(2)_L \times SU(2)_R$
admits a canonical real structure and the resulting four-dimensional real vector
space, as a representation of $SU(2) \times SU(2)$ can be identified with the
quaternions, as a representation of $U(1,\IH) \times U(1,\IH)$. The resulting
canonical basis is:
\be\label{eq:CanonBasis}
\begin{split}
\vert 1 \rangle & = \frac{1}{\sqrt{2}} ( \vert + -  \rangle - \vert - + \rangle ) \\
\vert 2 \rangle & = \frac{1}{\sqrt{2}} ( \vert + +  \rangle + \vert - - \rangle ) \\
\vert 3 \rangle & = \frac{\I}{\sqrt{2}} ( \vert + +  \rangle - \vert - - \rangle ) \\
\vert 4 \rangle & = \frac{\I }{\sqrt{2}} ( \vert + -  \rangle + \vert - + \rangle ) \\
\end{split}
\ee
This slightly peculiar basis of states appeared in \cite{gtvw}. It is determined by the quaternionic
structure and in particular is compatible with the real structure. Readers interested in
understanding the above remarks more thoroughly can consult Appendix \ref{app:DistBasis}.

The Pauli matrices $h(x)$ when acting under the diagonal embedding $\rho_{LR}(h(x)) = (h(x),h(x))$
will act diagonally on this basis, with entries $\pm 1$ along the diagonal. We will represent such a
matrix by a column vector with entries $0,1$. We convert $+ \leftrightarrow 0$
and $- \leftrightarrow 1$. Thus, for example, $(h(\uone), h(\uone))$ acts in this basis
as the diagonal matrix
\be
\begin{pmatrix}
+ & & &  \\
 & +  & &  \\
  &  & -  &  \\
   &  &   & -   \\
\end{pmatrix}
\ee
and we summarize that action by the column vector
\be
 \begin{pmatrix} 0 \\ 0 \\ 1 \\ 1 \\ \end{pmatrix}
\ee
In this way, the  signs appearing in the action of
$\rho_{LR}(h(x))$, for $x\in \IF_4$, on the canonical basis of $\IH \cong ( (1/2) \otimes (1/2))_{\IR}$
are neatly encoded in the map $g^+(x,0)$ defined in equation \eqref{eq:fg-LR-1-app}.

Next, when we act with the operators
$(h(w), h(w))$, for $w \in \IF_4^6$ on the distinguished basis of the RR sector we obtain a $4 \times 6$ array
of elements of $\IF_2$. This array can, in turn, be identified with a vector in $\IF_2^{24}$.

The action of $\widehat{Q}^6_{P}$ on $\CH_{GTVW}^{1/4, 1/4}$ factors through to an action of an Abelian group
\be
\left( \IF_4^{+} \times \IF_2\right)^6 \times \IZ_2
\ee
Acting on $\CH_{GTVW}^{1/4, 1/4}$ in the distinguished basis we find an action of the Golay code -
in the following sense:

Group elements of the form
\be
\left(  \begin{pmatrix} \epsilon_L^1  h(x_1) & 0  \\  0 &  \epsilon_R^1  h(x_1) \\ \end{pmatrix}, \dots,
\begin{pmatrix} \epsilon_L^6  h(x_6) & 0  \\  0 &  \epsilon_R^6  h(x_6) \\ \end{pmatrix} \right)
\ee
acts on the distinguished basis of RR states as:
\be
V^{i,\alpha} \to (-1)^{g^+(x_\alpha, \epsilon_L^\alpha \epsilon_R^\alpha)_i} V^{i,\alpha}
\ee
while group elements
\be
\left(  \begin{pmatrix} 0 & \epsilon_L^1  h(x_1)   \\   \epsilon_R^1  h(x_1) & 0 \\ \end{pmatrix}, \dots,
\begin{pmatrix}0 &  \epsilon_L^6  h(x_6)  \\    \epsilon_R^6  h(x_6) & 0  \\ \end{pmatrix}\right)
\ee
acts as
\be
V^{i,\alpha} \to (-1)^{g^-(x_\alpha, \epsilon_L^\alpha \epsilon_R^\alpha)_i  } V^{i,\alpha}
\ee
Comparing with the description of the Golay code in equations \eqref{eq:GolayEven1} to
\eqref{eq:odd-golay-words} we arrive at one of our main statements:

\emph{ Consider the stabilizer of $\Psi_L - \Psi_R$ within a left-right symmetric version of the
quaternion group (or group of error operators), namely the group
$\widehat{Q}^6_P$ defined above. This stabilizer group is a non-Abelian group which, when acting on $V_{RR}$
in the distinguished basis defined by the $SU(2)^6$ WZW model and the quaternions,  defines the Golay code.
 In equations, there is a natural isomorphism}
\be\label{eq:GolayFromN1}
\CG \cong \rho_{V_{RR}} \left( {\rm Stab}_{\widehat{Q}^6_{P} } (\Psi_L - \Psi_R) \right) ~ .
\ee

This gives a physical interpretation of the MOG presentation of the Golay code.
Note that, if we do not consider the extension by parity, and only consider the
left-right-symmetric action of the group preserving $\Psi$ then we obtain the
even Golay code $\CG^+$.

What is the significance of this result? It is well-known that the automorphism
group of the Golay code is $M_{24}$. This then, gives a new interpretation of an
$M_{24}$ ``symmetry'' within a K3 sigma model. We put ``symmetry'' in quotation marks
because it is an automorphism group of a symmetry group. It is not clear to us
what implication such a ``symmetry of a group of symmetries'' has for twined elliptic
genera and the space of massive BPS states. It is possible that the emerging ideas in the
context of ``generalized symmetries'' and domain walls will shed further light on this question.

There are two natural directions in which the above could
be extended: First, we have once again left off any investigation
of the role of the ``rotational'' symmetries $\Aut(\CH_6)$ in
the holomorph of the hexacode. Second, we limited discussion to the
left-right symmetric action on the canonical RR basis. One could
of course, also consider the same group elements acting either on
the right or on the left. The action will then be by signed
permutation matrices.

\section{Superconformal Symmetry And QEC In Conway Moonshine} \label{sec:Conway}

The methods of this paper shed some light on the Conway Moonshine module
studied in \cite{Duncan_super, Duncan:2014eha}  since very similar techniques can be
used to construct the $N=1$ supercurrent that plays a starring role in the analysis of those papers.
We are merely making concrete and explicit some points left implicit in \cite{Duncan_super, Duncan:2014eha}.

The Conway Moonshine module can be thought of as a theory of 24 Majorana-Weyl spinors
$\psi_i$, $i=1,\dots, 24$. This theory has a ${\rm Spin}(24)$ automorphism group.
 In the Ramond sector the ground states form a $2^{12}$ dimensional represention of ${\rm Spin}(24)$
 with a natural real structure.
The vertex operators associated to these states have conformal dimension
$h=24\times \frac{1}{16} = \frac{3}{2}$, and therefore there is a $2^{12}$ dimensional
space of potential supercurrents $V_s$ labeled by ${\rm Spin}(24)$ spinors $s$.
Using the representation theory of the Conway group
\cite{Duncan_super, Duncan:2014eha} showed that there is a distinguished
spinor $s$ such that $V_s$ indeed defines a superconformal current,
and moreover, the stabilizer of $Spin(24)$ acting on $s$ is exactly
the largest sporadic Conway group, thus identifying the Conway group as a group of $N=1$
supersymmetry preserving automorphisms. What we will do here is construct
the spinor explicitly using the Golay code and show, in an elementary
way, how the properties of the Golay code imply the required identities for $V_s$
to define a supercurrent.

The OPE of $V_s$ with itself is given by  \cite{Witten:2007kt}
\be
V_s(1) V_s(2) \sim \frac{s^{tr}s}{z_{12}^3} + \frac{ s^{tr}\gamma^{ij} s }{z_{12}^2} \psi_i \psi_j
+ \frac{s^{tr} s}{z_{12}} T + \frac{s^{tr} \gamma^{ijkl} s}{z_{12}} \psi_i \psi_j \psi_k \psi_l + \cdots
\ee
This follows from ${\rm Spin}(24)$ symmetry.
Therefore, the  OPE of $V_s$ with itself will define a supercurrent provided
\be\label{eq:firstcond}
s^{tr}  \gamma^{ij} s  = 0 \qquad\qquad  i<j
\ee
\be\label{eq:secondcond}
s^{tr}  \gamma^{ijkl} s  = 0   \qquad\qquad  i<j<k<l
\ee

We will now construct a solution to these equations. In the Ramond sector the
zeromodes of $\psi_i$  will be $24$
Clifford generators $\gamma_i$ with $\gamma_i^2=1$. Now, for $w \in \IF_2^{24}$ let
\be
\gamma_w:= \gamma_1^{w_1} \cdots \gamma_{24}^{w_{24}}
\ee
Then
\be
\gamma_{w_1} \gamma_{w_2} = \epsilon(w_1, w_2) \gamma_{w_1+w_2}
\ee
where
\be
\epsilon(w_1, w_2) = (-1)^{ \sum_{i<j} w_2^i w_1^j }
\ee
is a nontrivial cocycle. Indeed, it describes the Heisenberg extension of $\IF_2^{24}$ by $\IF_2$.
The key observation is that, when restricted to the Golay code $\CG\subset \IF_2^{24}$ the
cocycle is trivializable by a $\pm 1$-valued coboundary.

We will use the MOG presentation to represent Golay code words as
$\mathfrak{f}^+(x,\epsilon )$ or $\mathfrak{f}^-(x,\epsilon)$ where $(x,\epsilon)$ is a decorated hexacode
word as described between equations \eqref{eq:GolayEven1} to \eqref{eq:odd-golay-words}.
Label the basis vectors and gamma matrices by   $1,2,3,4$ going from top to bottom
in the first column, $5,6,7,8$ top to bottom in the second column, etc.

Then, we let, for example:
\be
\gamma_{g^+(\uz,0)} = 1
\ee
\be
\gamma_{g^+(\uz,1)} = \gamma_1\gamma_2\gamma_3 \gamma_4 = \gamma_{1234}
\ee
\be
\gamma_{g^+(\uone,0)} = \gamma_3 \gamma_4 = \gamma_{34}
\ee
\be
\gamma_{g^+(\uone,1)} = \gamma_1 \gamma_2 = \gamma_{12}
\ee
\be
\gamma_{g^+(\uw,0)} = \gamma_2 \gamma_4 = \gamma_{24}
\ee
\be
\gamma_{g^+(\uwb,0)} = \gamma_2 \gamma_2 = \gamma_{23}
\ee

Now, if $w=((x_1,\epsilon_1), \dots , (x_6, \epsilon_6))=(x,\epsilon) $ is a  decorated hexacode word we let
\be
\gamma_{\ff^+(w) } := \gamma_{g+(x_1,\epsilon_1) }^{(1)} \cdots \gamma_{g+(x_6, \epsilon_6) }^{(6)}
\ee
where the superscript in parentheses indicates which quartet of gamma matrices
we are using.   For $\gamma_{g+(x_1,\epsilon_1)}^{(1)}$ we use
the set of four $\gamma$-matrices $\gamma_1,\dots, \gamma_4$ for $\gamma_{g+(x_2,\epsilon_2) }^{(2)} $ we use
the set of four $\gamma$-matrices $\gamma_5,\dots, \gamma_8$, and so on.

A computation shows that if $w\in \CG$ corresponds to $(x,\epsilon)$ and we define
\be\label{eq:Trivialize}
b(w) = (-1)^{\sum_{\alpha} \delta(x_\alpha) \epsilon_\alpha}
\ee
where, for $x_\alpha \in \IF_4$,
\be
\delta(x_\alpha) := \begin{cases} 0 & x_\alpha=\uz,\uw  \\   1 & x_\alpha = \uone,\uwb \\
\end{cases}
\ee
then, for $w,w' \in \CG$ we have
\be
\gamma_w \gamma_{w'} = \frac{b(w+w')}{b(w) b(w')} \gamma_{w+w'}
\ee
It therefore follows that if we define $\tilde \gamma_w: = b(w) \gamma_w$ then
\be
P_D := \frac{1}{2^{12}} \sum_{w\in \CG} \tilde \gamma_w
\ee
is a rank one projection operator. Duncan's spinor is in the image of this projection
operator: $s_{D} \propto P s_0$ for any generic spinor $s_0$.

It is now elementary to explain why equations \eqref{eq:firstcond} and
\eqref{eq:secondcond} are satisfied by Duncan's spinor. We observe that 
\be 
s_D s_D^{tr} = k P_D 
\ee
for a suitable constant $k$. (We are working in a real vector 
space $\IR^{2^{12}}$ with Euclidean norm.) Therefore 
\be 
\begin{split} 
s_D^{tr} \gamma_{ij} s_D & = {\rm Tr} \gamma_{ij} s_D s_D^{tr} \\ 
& =k  {\rm Tr} \gamma_{ij} P_D  \\ 
& = \frac{k}{2^{12}} \sum_{w \in \CG} b(w) {\rm Tr} \gamma_{ij} \gamma_w  \\ 
\end{split}
\ee
and similarly for $s_D^{tr} \gamma_{ijkl} s_D$.
Next note that
\be
\gamma_{ij}   \gamma_w = \pm \gamma_{w + e_i + e_j  }
\ee
\be
 \gamma_{ijkl} \gamma_w = \pm \gamma_{w + e_i + e_j + e_k + e_l }
\ee
but, thanks to the error-correcting properties of the Golay code, this means
that $w+e_i + e_j$ and $w + e_i + e_j + e_k + e_l$ are never Golay code words.
Thus, $ {\rm Tr} \gamma_{ij} \gamma_w  =0$ for all Golay code words $w$ and $i<j$, 
and similarly  $ {\rm Tr} \gamma_{ijkl} \gamma_w  =0$.
The proof is thus closely analogous to our proof that $V_\Psi$ generates a
superconformal current in the GTVW model.

An argument closely analogous to that of section \ref{subsec:FiniteStab}
shows that the stabilizer group of the image of $P$ is a finite subgroup of ${\rm Spin}(24)$.
It is easy to show from the above description of the projection operator
that the stabilizer of Duncan's spinor contains the maximal
subgroup $2^{12}:M_{24}$ of $Co_0$. The fact that the stabilizer
is exactly $Co_0$ is more nontrivial and follows from general arguments in
\cite{Duncan_super,Duncan:2014eha}. It would be very nice to demonstrate this
directly using the above description for the projection operator. Such a
demonstration appears to be nontrivial, and this will be left for another
occasion.

\subsection{Relation Between Conway And GTVW Superconformal Currents}

There is a close relation between the, so-called ``reflected GTVW model'' and the
Conway Moonshine module explored in \cite{Creutzig:2017fuk, Taormina:2017zlm}.
Starting with the Conway Moonshine module, we
split the 24 fermions according to the MOG: $\psi_{\alpha}^a$, with $a=1,\dots, 4$.
For each $\alpha$ we have a  $\fs\fo(4) \cong \fs\fu(2)_{\alpha}^L \oplus \fs\fu(2)_{\alpha}^R $
$k=(1,1)$ current algebra. We then have six copies of the VOA's these generate corresponding to
 the six columns of the MOG
and we identify that as a subalgebra of the reflected GTVW theory. Now the reflected
GTVW theory has an $N=1$ superconformal current
\footnote{Indeed, one take the reflected sum of each of the four supercurrents
leading to an $N=4$ structure on the reflected GTVW model \cite{Taormina:2017zlm}.}
\be
V_{\Psi_L} + V_{\Psi_R}
\ee
with $c=12$ and energy-momentum tensor
\be
T = - \half \psi_{\alpha}^a \p \psi_{\alpha}^a
\ee
Given the uniqueness of the $N=1$ structure $V_{\Psi_L} + V_{\Psi_R} $
should coincide with   $V_{s_{Duncan}}$ up to automorphism of the CFT.
We can see this rather nicely as follows: For any pair $(\alpha_1, \alpha_2)$
of columns in the MOG consider the decorated hexacode word $(x,\epsilon)$ with
$x_{\alpha} = \uz$ for all $\alpha$ and
\be
\epsilon_\alpha = \begin{cases}
0 & \alpha \not= \alpha_1, \alpha_2 \\
1 & \alpha \in \{ \alpha_1, \alpha_2 \} \\
\end{cases}
\ee
Then $\ff^+(x,\epsilon)$ is an even interpretation of the zero hexacode word,
the trivialization $b(x,\epsilon)=1$ and
\be
\tilde \gamma_{\ff^+(x,\epsilon)}
\ee
is just the product of the column chirality operators for columns $\alpha_1$ and $\alpha_2$.
Thus, $s_{D}$ can be expressed as $s^+ + s^-$ where $s^+$ has all column chiralities
equal to $+1$ and $s^-$ has all column chiralities equal to $-1$.

Now note that if  we decompose the
spinor representation of $\fs\fo(24)$ under the subalgebra
\be
\oplus_{\alpha} \fs\fo(4)_{\alpha}
\ee
then we get:
\be
\left( (\textbf{2} ;0) \oplus \widetilde{ (0;\textbf{2} )} \right)_{\alpha=1} \otimes \cdots
\left( (\textbf{2};0) \oplus \widetilde{ (0;\textbf{2})} \right)_{\alpha=6}  \cong
\oplus_{x + \tilde x = e}  V_x \otimes \tilde V_{\tilde x}
\ee
where $V_0$ is the singlet of $\fs\fu(2)$ and $V_1$ is the doublet of $\fs\fu(2)$
and $e$ is the all ones vector.
The GTVW supercurents live in the components
\be
\begin{split}
V_{\Psi_L} & \in  V_{1^6} \otimes \tilde V_{0^6} \\
V_{\Psi_R} & \in  V_{0^6}\otimes \tilde V_{1^6}\\
\end{split}
\ee
but $V_{1^6} \otimes \tilde V_{0^6}$ is the image of the
 projection operators:
\be
P_+ = \left( \frac{ 1 + \gamma^{1234} }{2} \right)_{\alpha=1} \otimes \cdots \otimes
\left( \frac{ 1 + \gamma^{1234} }{2} \right)_{\alpha=6}
\ee
onto the space of spinors with all column chiralities $=+1$ and
$ V_{0^6}\otimes \tilde V_{1^6}$ is the image of the projection
operator
\be
P_- = \left( \frac{ 1 - \gamma^{1234} }{2} \right)_{\alpha=1} \otimes \cdots \otimes
\left( \frac{ 1 - \gamma^{1234} }{2} \right)_{\alpha=6}
\ee
with all column chiralities $=-1$. This shows that
\be
\begin{split}
V_{\Psi_L} & = s^+ \\
V_{\Psi_R} & = s^- \\
\end{split}
\ee
We can write:
\be
P_{D} = 2^{-12} (\sum_{w\in \CG^+} \tilde \gamma_{\ff^+} +  \sum_{w\in \CG^-} \tilde \gamma_{\ff^-})
\ee
The first term preserves all column chiralities and the second term changes all column chiralities.
The first term can be restricted to $V_{1^6}\otimes \tilde V_{0^6}$, and this should coincide
with $P_{GTVW}$.

%
%
%
%
%

\appendix

\section{The Automorphism Group Of The Hexacode And The Even Golay Code}\label{app:HexacodeAutGroup}

In this paper we adopt notation for finite groups and their extensions used in \cite{atlas}. In particular, $p^m$ indicates
the group $(\IZ/p \IZ)^m$, $A \times B$ is the direct product of the groups $A$ and $B$, $A.B$ indicates a group with normal subgroup $A$
and quotient isomorphic to $B$ while
 $A:B$ denotes a group which is a semi-direct product of $A$ and $B$. Of course for the latter a full description requires specifying a homomorphism $\phi: B \rightarrow {\rm Aut}(A)$.

For any group $G$ we define the \emph{holomorph of $G$} to be the
group ${\rm Hol}(G):=G : \Aut(G)$ where the semi-direct product is
defined using the natural action of $\Aut(G)$ on $G$. Thus, the
group ${\rm Hol}(G)$ acts naturally on $G$ itself
where the first factor acts by (say) left-translation. A good example is the
group of automorphisms of real $n$-dimensional affine space,
which is isomorphic to the holomorph of the group $\IR^n$.

The hexacode has some very useful symmetries, and in this
appendix we review the structure of it automorphism group
$\Aut(\CH_6)$ in some detail.

First of all,   being a linear subspace of $\IF_4^6$,
a hexacode word is mapped to another by multiplication
by any scalar, and if the scalar is nonzero this is an
automorphism. We denote the group of scalar multiplication by
nonzero elements of $\IF_4$ by $H_0$. Of course $H_0 \cong \IZ_3$
and one generator would be
\be
g_0: (x_1, \dots, x_6) \rightarrow (\uw x_1, \dots , \uw x_6)
\ee

Next, there are some simple permutation symmetries, i.e. subgroups
of the natural $S_6$ action on $\IF_4^6$ that preserve $\CH_6$.
To describe these   it is useful to arrange a 6-digit word in $\CH_6$
as 3 couples: $(x_1, \dots, x_6) = (ab~ cd~ef )$.
One such group of symmetries is obtained by flipping {\it pairs} of
couples. We can take as generators:
\be
g_1 = (12)(34)
\ee
\be
g_2 = (34)(56)
\ee
these generate a group  $H_1 \cong  \IZ_2 \times \IZ_2$. Another subgroup of permutation symmetries is obtained
by arbitrary permutation of couples. This defines a subgroup $H_2 \cong S_3$. One choice of generators of $H_2$
would be:
\be
g_3 = (13)(24)
\ee
\be
g_4 = (35)(46)
\ee
Clearly $H_2$ normalizes $H_1$ and together
these generate a group $H_3 = H_1 : H_2 \cong S_4$.
One way to prove that $H_3$ is a group of symmetries of the hexacode proceeds by
applying the generators $g_1, \dots, g_4$ to
three basis vectors in equation \eqref{eq:hexbasis} and checking that the resulting vectors remain in the hexacode.
Together with scalar multiplication we obtain a subgroup $H_0 \times H_3 \cong \IZ_3 \times S_4$ of the automorphism
group. One can check directly that the  orbit of the four ``seed codewords''
\be\label{eq:SeedWords}
 (\uone \uone ~\uw \uw ~\uwb \uwb )
\qquad (\uz \uz ~ \uone \uone ~ \uone \uone) \qquad (\uw \uwb ~ \uw \uwb ~ \uw \uwb )
\qquad (\uz \uone ~ \uz \uone ~ \uw \uwb)
\ee
under $H_0 \times H_3$ is the entire set of nonzero words in the hexacode.
See eqn. \ref{eqn:Hexacode}.

There are further, ``nonobvious'' automorphisms of the hexacode.
An example of such an automorphism is
\be
g_5: (x_1, \dots, x_6) \rightarrow (\uw x_1 , \uwb x_2, x_3, x_6 , x_4, x_5)
\ee
To prove that $g_5$ is a symmetry note that $(x_1, \dots, x_6) \in \CH_6$ iff
\be
\begin{split}
x_4 & = \Phi_{x_1, x_2, x_3}(\uone) = x_1 + x_2 + x_3 \\
x_5 & = \Phi_{x_1, x_2, x_3}(\uw) = \uwb x_1 + \uw x_2 + x_3 \\
 x_6 & = \Phi_{x_1, x_2, x_3}(\uwb) = \uw x_1 + \uwb x_2 + x_3 \\
\end{split}
\ee
(See equation \eqref{eq:quadfundef} above.)
Then, letting $g_5\cdot (x_1, \dots, x_6)= (y_1, \dots y_6)$,
\be
\begin{split}
y_4 & = \Phi_{y_1, y_2, y_3}(\uone) = y_1 + y_2 + y_3 = \uw x_1 + \uwb x_2 + x_3 = x_6 \\
y_5 & = \Phi_{y_1, y_2, y_3}(\uw) = \uwb y_1 + \uw y_2 +   y_3 = x_1 + x_2 + x_3 = x_4 \\
y_6 & = \Phi_{y_1, y_2, y_3}(\uwb) = \uw y_1 + \uwb y_2 + y_3 = \uwb x_1 + \uw x_2 + x_3 = x_5 \\
\end{split}
\ee
Let $H_5:= \langle g_1, \dots, g_5 \rangle$.
There is a projection
\be\label{eq:TrackPerm}
p: \Aut(\CH_6)\to S_6
\ee
where the image of $p$ just tracks how the automorphism permutes the
hexacode digits. Clearly, the kernel of $p$ is just $H_0$. Moreover,
one can check by direct computation
that the image of $H_5$ under $p$ is the entire subgroup $A_6 \subset S_6$.
Thus, $H_{05}:=\langle g_0, \dots, g_5 \rangle$ is a central extension of $A_6$
by $H_0 \cong \IZ_3$. By computing the lift of two elements in $A_6$ whose
group commutator vanishes one easily checks that it is a nontrivial
central extension, so
\footnote{For a nice discussion see Lecture one of \cite{Theolectures}.}
\be\label{eq:H05-def}
H_{05}:=\langle g_0, \dots, g_5 \rangle   \cong \IZ_3 \cdot A_6
\ee

Another example of a non-obvious automorphism of $\CH_6$ is
\be
g_F: (x_1, \dots, x_6) = (56) \cdot (x_1^2, \dots, x_6^2) = (56) \cdot ( \bar x_1 , \dots, \bar x_6 )
\ee
In the second equality we have used the fact that the nonlinear map $x \to x^2$ is identical to the
Frobenius automorphism $x\to \bar x$ of $\IF_4$. To prove that $g_F$ is an automorphism of the hexacode we again
use equation \eqref{eq:quadfundef} and the result follows immediately. Since $p(g_F) = (56)$ is an odd permutation it is clear that
the image under $p$ of $\langle g_1, \dots, g_5, g_F \rangle$ is all of $S_6$.
On the other hand, $g_F$ does not commute with $g_0$ and conjugation by $g_F$
acts as the nontrivial automorphism of $H_0$. Thus,
\be
\langle g_0, g_1, \dots, g_5, g_F \rangle \cong H_0 \cdot  \langle g_1, \dots, g_5 , g_F \rangle \cong \IZ_3 \cdot S_6.
\ee
In \cite{Wilson} it is asserted that the full automorphism group $\Aut(\CH_6) \cong \IZ_3 \cdot S_6$ so we have
now described in detail the full structure of the automorphism group of the hexacode.

In computations it can be useful to have a full list of hexacode words. We provide this list in eqn. \ref{eqn:Hexacode} which
contains all hexacode words with the exception of the trivial word $00~00~00$. The words
are organized into orbits of $H_2 \cong S_3$ and  $H_1 \cong \IZ_2 \times \IZ_2$ generated by $g_1, g_2$.
\begin{align} \label{eqn:Hexacode}
\begin{split}
\substack{S_3  {~\rm orbit}, \\ g_1, g_2 {~\rm invariant} }
& \begin{cases}
 & \rightarrow 11~\om \om~\omb \omb \rightarrow \om \om~\omb \omb~ 1 1 \rightarrow \omb \omb~1 1~\om \om \rightarrow \\
& \rightarrow \om \om~11~\omb \omb \rightarrow \omb \omb~\om \om~ 11 \rightarrow 11~ \omb \omb~\om \om \rightarrow \\
\end{cases}  \\
\substack{S_3 ~{\rm orbit},  \\ g_1, g_2 ~{\rm exchanges~rows} }
& \begin{cases}
& \rightarrow \om \omb ~\om \omb~ \om \omb \rightarrow \omb 1~ \omb 1~ \omb 1 \rightarrow 1 \om~1 \om~ 1 \om \rightarrow \\
& \rightarrow \omb \om ~ \omb \om ~ \om \omb \rightarrow 1 \omb ~ 1 \omb ~ \omb 1 \rightarrow \om 1 ~ \om1 ~ 1 \om \rightarrow \\
& \rightarrow  \omb \om~\om \omb~\omb \om\rightarrow 1 \omb~ \omb 1~ 1 \omb \rightarrow \om 1~ 1 \om ~ \om 1 \rightarrow \\
&\rightarrow \om \omb~ \omb \om ~ \omb \om \rightarrow  \omb 1 ~ 1 \omb~ 1 \omb \rightarrow 1 \om~ \om 1~ \om 1 \rightarrow
\end{cases} \\
\substack{ S_3 ~{\rm orbit}, \\ g_1, g_2 ~{\rm invariant}  }
& \begin{cases}
 & \rightarrow 00~11~11 \rightarrow 00~\om \om~ \om  \om \rightarrow 00~\omb \omb~\omb \omb \rightarrow \\
& \rightarrow 11~00~11 \rightarrow \om \om~00~\om \om \rightarrow \omb \omb~ 00~\omb \omb \rightarrow \\
& \rightarrow 11~11~00 \rightarrow \om \om~ \om \om~00 \rightarrow \omb \omb~\omb \omb~00 \rightarrow \\
\end{cases} \\
\substack{ S_3 ~{\rm orbit}, \\ {\rm rows}~1-4, 5-8, 9-12 \\ ~{\rm are~each} \\ g_1, g_2 ~{\rm orbits}  }
& \begin{cases}
& \rightarrow 0 1~0 1~ \om \omb \rightarrow 0 \om~0 \om ~ \omb 1 \rightarrow 0 \omb ~0 \omb~ 1 \om \rightarrow \\
& \rightarrow 1 0 ~ 1 0 ~ \om \omb \rightarrow \om 0 ~ \om 0 ~ \omb 1 \rightarrow \omb 0 ~\omb 0 ~ 1 \om \rightarrow \\
&\rightarrow 0 1 ~ 1 0 ~ \omb \om \rightarrow 0 \om ~ \om 0 ~ 1 \omb \rightarrow 0 \omb ~ \omb 0 ~ \om 1 \rightarrow \\
& \rightarrow 1 0 ~ 0 1 ~ \omb \om \rightarrow \om 0 ~ 0 \om ~ 1 \omb \rightarrow \omb 0 ~ 0 \omb ~ \om 1 \rightarrow \\
& \rightarrow 0 1  ~ \om \omb   ~ 0 1 \rightarrow 0 \om ~ \omb 1 ~  0 \om \rightarrow 0 \omb~ 1 \om ~ 0 \omb  \rightarrow \\
& \rightarrow 1 0  ~\omb \om   ~ 0 1 \rightarrow \om 0 ~  1 \omb ~ 0 \om \rightarrow \omb 0~  \om 1~ 0 \omb \rightarrow  \\
& \rightarrow 0 1  ~ \omb \om   ~ 1 0 \rightarrow 0 \om~ 1 \omb ~ \om 0 \rightarrow 0 \omb~ \om 1~ \omb 0 \rightarrow \\
&  \rightarrow  1 0 ~ \om \omb   ~ 1 0 \rightarrow  \om 0~ \omb 1~  \om 0 \rightarrow \omb 0 ~ 1 \om ~ \omb 0  \rightarrow  \\
&  \rightarrow  \om \omb  ~ 0 1   ~ 0 1 \rightarrow  \omb 1~  0 \om~ 0 \om  \rightarrow  1\om ~ 0 \omb ~ 0 \omb \rightarrow  \\
& \rightarrow   \omb \om ~ 1 0  ~ 0 1 \rightarrow 1 \omb ~ \om 0 ~ 0 \om  \rightarrow  \om 1 ~\omb 0 ~ 0 \omb \rightarrow  \\
& \rightarrow  \om \omb  ~ 1 0  ~ 1 0 \rightarrow \omb 1 ~ \om 0~  \om 0 \rightarrow 1 \om ~ \omb 0 ~  \omb 0 \rightarrow \\
&  \rightarrow  \omb \om ~  0 1  ~ 1 0 \rightarrow 1 \omb ~ 0 \om~ \om 0 \rightarrow \om 1 ~ 0 \omb ~ \omb 0  \rightarrow
\end{cases}   \\
\end{split}
\end{align}

\bigskip
\noindent
\textbf{Remark}: It is quite interesting to note that the automorphisms
of the hexacode act on two distinct sets of six objects. The first is
the set of six digits. However, the automorphisms also permutes the no-zeroes
hexacode words (i.e. those words where none of the digits is zero). Up
to overall scale there are exactly six types of no-zeroes hexacode words.
They correspond to the first six lines of \eqref{eqn:Hexacode}. The relation
between these two group actions of $S_6$ on sets with six elements is related
by an exceptional outer automorphism of the symmetric group $S_6$.

\section{Supersymmetry Conventions}\label{app:SuperAlgebras}

\subsection{Superconformal Algebras}

The $N=1$ 2d superconformal algebra has generators $G_r, L_m$ with
commutation relations
\be
\begin{split}
\{ G_r, G_s \} & = 2L_{r+s} + \frac{c}{12} (4r^2 - 1) \delta_{r+s,0} \\
[L_m, G_r] & = \left( \frac{m}{2}-r\right) G_{m+r } \\
[L_m, L_n]  & = (m-n) L_{m+n} + \frac{c}{12} (m^3- m) \delta_{m+n,0} \\
\end{split}
\ee
Here $m \in \IZ$ and $r\in \IZ$ or $r \in \IZ + \half$ for the Ramond or Neveu-Schwarz algebra respectively. In terms of OPE's of the currents:
\be
\begin{split}
T(z) & = \sum_n z^{-n-2} L_n \\
G(z) & = \sum_r z^{-r-3/2} G_r \\
\end{split}
\ee
we have
\be\label{eq:N1-SCA-OPE}
\begin{split}
G(z) G(w) & \sim \frac{\frac{  2c}{3} }{(z-w)^3}  + \frac{\half T(w)}{z-w} + \cdots \\
T(z) G(w) &  \sim \frac{ \frac{3}{2} G(w)}{(z-w)^2} + \frac{ \p G(w)}{z-w} + \cdots \\
T(z) T(w) & \sim \frac{ \frac{c}{2} }{(z-w)^4} + \frac{2 T(w) }{(z-w)^2} + \frac{ \p T(w) }{z-w} + \cdots \\
\end{split}
\ee

The small $N=4$ superconformal algebra has generators $L_n, Q^a_r, T_m^i$ with relations:
\be
\begin{split}
[L_m,L_n] & = (m-n) L_{m+n} + \frac{k}{2}(m^3-m) \delta_{m+n,0} \\
\{ Q^a_r, Q^b_s\} & = \{ \bar Q^a_r, \bar Q^b_s\} = 0 \\
\{ Q^a_r, \bar Q^b_s\}  & = 2 \delta^{ab} L_{r+s} - 2 (r-s) \sigma^i_{ab} T^i_{r+s} + \frac{k}{2}(4r^2-1)\delta_{r+s,0} \delta^{ab} \\
[T^i_m, T^j_n] & = i \epsilon^{ijk} T^k_{m+n} + \frac{k}{2} m \delta_{m+n,0} \delta^{ij} \\
[T_m^i , Q^a_r]  = - \half \sigma^i_{ab} Q^b_{m+r} \qquad & \qquad [T_m^i ,\bar Q^a_r]  =  \half (\sigma^i_{ab})^* \bar Q^b_{m+r} \\
[L_m , Q^a_r]  = \left( \frac{m}{2}-r \right) Q^a_{m+r} \qquad & \qquad [L_m ,\bar Q^a_r]  =  \left( \frac{m}{2}-r \right) \bar Q^a_{m+r} \\\
[L_m, T^i_{n}] & = -n T^i_{m+n} \\
\end{split}
\ee
 Here it is traditional to parametrize by $c=6k$ and $k \in \IZ_+$ for unitary theories.
Here we are using the conventions of \cite{Eguchi:1988af,Eguchi:1988vra}. There is a natural real structure on this algebra
defined by
\be\label{eq:N4-realstructure}
(Q^a_r)^\dagger = \bar Q^a_{-r} \qquad\qquad (T_n^i)^\dagger = T_{-n}^i
\ee
Note that since the $T_0^i$ are real they will be represented by Hermitian operators in a
unitary representation  whereas
in our conventions   $J^i_0$ are antihermitian.

%
%

As explained above we are interested in embeddings of the $N=1$ superconformal
algebra into the $N=4$ algebra. We thus require that a linear combination of
$N=4$ supercurrents
\be
G_r = \alpha_1 Q^1_r + \alpha_2 Q^2_r + \beta_1 \bar Q^1_r + \beta_2 \bar Q^2_r
\ee
satisfies the $N=1$ superconformal algebra. Moreover,  we require that under
the real structure \eqref{eq:N4-realstructure} we have $(G_r)^\dagger = G_{-r}$.
It is not difficult to show that, up to an overall $SU(2)_R$ rotation the most general solution
is
\be
G = \half (Q^1 + \bar Q^1)
\ee
Note that this linear combination is {\it not} an eigenstate of $T^3_0$. In fact, this is a vector in the
$\textbf{2} \oplus \textbf{2}$ that completely breaks continuous $SU(2)_R$-symmetry.

To make contact with the notation used in \cite{gtvw} we note that
\be
Q^1 \propto G^+ \qquad \bar Q^1  \propto G^- \qquad \bar Q^2  \propto G^{'+} \qquad Q^2  \propto G^{' -}
\ee
Since   $(G^+, G^{',-})$ is an $SU(2)_R$ doublet and $(G^{',+}, G^-)$ is another $SU(2)_R$ doublet.

\section{Proof That There Are No Nonzero Solutions Of \eqref{eq:Inf-Stab}}\label{app:ProofDiscrete}

To show that \eqref{eq:Inf-Stab} has no nontrivial solutions we write
\be
\CO := \sum_{x\in \IF_4^*} \sum_{\alpha=1}^6  c_{x,\alpha} h(x)_\alpha
\ee
As a sum
\be
\CO = \CO_{diag} + \CO_{flip}
\ee
\be
\CO_{diag} := \sum_{\alpha=1}^6  c_{\uwb,\alpha} h(\uwb)_\alpha
\ee
\be
\CO_{flip} := \sum_{\alpha=1}^6 ( c_{\uone,\alpha} h(\uone)_\alpha + c_{\uw,\alpha}  h(\uw)_\alpha )
\ee
The idea here is that $\CO_{diag}$ acts on the spin states in $\Psi$ just by multiplying
by a phase, while $\CO_{flip}$ flips one digit. So these terms cannot interfere.

So, setting $\CO_{diag} \Psi =0$ we get 16 equations. For example, acting on $[\emptyset]$ we get
\be
-\I (c_{\uwb,1} + c_{\uwb,2} + \cdots + c_{\uwb,6}) = 0
\ee
while acting on $[12]$ we get:
\be
\I (c_{\uwb,1} + c_{\uwb,2}) -\I (c_{\uwb,3} + c_{\uwb,4} + c_{\uwb,5} + c_{\uwb,6}) =0
\ee
Altogether we 16 equations. They are not all independent but we find independent equations
\be
\begin{split}
c_{\uwb,1} + c_{\uwb,2} + \cdots + c_{\uwb,6} & = 0 \\
(c_{\uwb,1} + c_{\uwb,2}) - (c_{\uwb,3} + c_{\uwb,4} + c_{\uwb,5} + c_{\uwb,6}) & = 0 \\
(c_{\uwb,3} + c_{\uwb,4}) - (c_{\uwb,1} + c_{\uwb,2} + c_{\uwb,5} + c_{\uwb,6}) & = 0 \\
(c_{\uwb,5} + c_{\uwb,6}) - (c_{\uwb,1} + c_{\uwb,2} + c_{\uwb,3} + c_{\uwb,4})& = 0 \\
(c_{\uwb,1} + c_{\uwb,3} + c_{\uwb,5} ) - (c_{\uwb,2} + c_{\uwb,4} + c_{\uwb,6}) & = 0 \\
(c_{\uwb,2} + c_{\uwb,4} + c_{\uwb,5} ) - (c_{\uwb,1} + c_{\uwb,3} + c_{\uwb,6}) & = 0 \\
(c_{\uwb,1} + c_{\uwb,4} + c_{\uwb,6} ) - (c_{\uwb,2} + c_{\uwb,3} + c_{\uwb,5}) & = 0 \\
\end{split}
\ee

Some of the tedium of writing these equations can be reduced by recalling that $\Psi$ is
symmetric under permutations of the couples $(12),(34),(56)$, and this symmetry must be
reflected in the equations. In any case, by adding and subtracting equations we quickly
find that $c_{\uwb,6} = - c_{\uwb,5}$, $c_{\uwb,4} = - c_{\uwb,3}$ and $c_{\uwb,2} = - c_{\uwb,1}$
and
\be
\begin{split}
c_{\uwb,1} + c_{\uwb,3} + c_{\uwb,5}  & = 0 \\
c_{\uwb,2} + c_{\uwb,4} + c_{\uwb,5}   & = 0 \\
c_{\uwb,1} + c_{\uwb,4} + c_{\uwb,6}    & = 0 \\
\end{split}
\ee
The determinant of the $3\times 3$ matrix is nonzero and hence $c_{\uwb,a}=0$.

Similarly, with the bit-flip operator $\CO_{flip}$ note that $[\emptyset]$ is mapped
to states $[a]$ with just one minus sign. While states like $[12]$ are mapped to
states with just one or just three minus signs. The equations split nicely. For example
the coefficients of the following output spin states give us the following equations:
\be
\begin{split}
[1] \qquad & \qquad - c_{\uone,1} + \I c_{\uw,1} + \I c_{\uone,2} - c_{\uw,2} = 0 \\
[2] \qquad & \qquad - c_{\uone,2} + \I c_{\uw,2} + \I c_{\uone,1} - c_{\uw,1} = 0 \\
[156] \qquad & \qquad -\I c_{\uone,1} - c_{\uw,1} +  c_{\uone,2} +\I c_{\uw,2} = 0 \\
[256] \qquad & \qquad -\I c_{\uone,2} - c_{\uw,2} +  c_{\uone,1} +\I c_{\uw,1} = 0 \\
\end{split}
\ee
Computing the determinant of the relevant $4\times 4$ matrix we find it is nonzero and hence
\be
 c_{\uone,1} =  c_{\uw,1} =  c_{\uone,2}= c_{\uw,2} = 0
\ee
Taking into account the symmetry of the permutations of couples, we conclude that
the identical equations hold for the couple $(34)$ and for $(56)$ and hence all the
$c_{x,a}=0$. It follows that the stabilizer group is a discrete subgroup of
$SU(2)^6$.

\section{Quaternions Give A Distinguished Basis For The $(\textbf{2}; \textbf{2})$ Representation Of $SU(2)\times SU(2)$}\label{app:DistBasis}

Let $U(\IH)$ denote the group of unit quaternions. There is a standard representation:
\bel
\xymatrix{
T: U(\IH) \times U(\IH) \ar[r] & GL(\IH)
}
\ee
defined by
\bel
T(q_1, q_2): q_0 \rightarrow q_1 q_0 q_2^{-1}
\ee
Here $GL(\IH)$ is the group of invertible linear transformations
regarding $\IH$ as a real vector space.

Now let $\qi, \qj, \qk$ be unit quaternions, and use the choice of matrices $h(x)$
in this paper to define a map
\bel
\IH \rightarrow {\rm Mat}_{2\times 2}(\IC)
\ee
by taking
\bel
\begin{split}
1 & \rightarrow h(\underline{0}) := \begin{pmatrix} 1 & 0 \\ 0 & 1 \\ \end{pmatrix} \\
\qi & \rightarrow h(\underline{1}) := \begin{pmatrix} 0 & 1 \\ -1 & 0 \\ \end{pmatrix} \\
\qj & \rightarrow h(\underline{\omega}) := \begin{pmatrix} 0 & \I \\ \I  & 0 \\ \end{pmatrix} \\
\qk & \rightarrow h(\underline{\bar \omega}) := \begin{pmatrix} -\I & 0 \\ 0 & \I \\ \end{pmatrix} \\
\end{split}
\ee
and extending $\IR$-linearly. Restricted to $U(\IH)$ this defines a group isomorphism:
\bel
\xymatrix{ U(\IH) \ar[r]^\psi & SU(2)}
\ee

On the other hand, we can   define a real representation of $SU(2)\times SU(2)$
\bel
\xymatrix{ R: SU(2) \times SU(2) \ar[r] & GL(V) }
\ee
where $V$ is a real four-dimensional vector space and $GL(V)$ is the group of real invertible
linear transformations on $V$.  This is just the $(2;2)$ representation with
a reality condition. Explicitly, letting $\alpha, \dot \beta $ run over $\{+,-\}$ in that order
vectors in $V$ can be written as
\bel
X_{\alpha, \dot \beta} \vert \alpha, \dot \beta \rangle
\ee
where we impose the reality condition:
\bel
(X_{\alpha\dot\beta})^* = \epsilon^{\alpha\alpha'} \epsilon^{\dot \beta \dot \beta'} X_{\alpha' , \dot \beta'}
\ee
The general solution of this reality condition is:
\bel
X_{\alpha\dot\beta} = \begin{pmatrix} z & - \bar w \\ w & \bar z \\ \end{pmatrix}
\ee
where $z,w \in \IC$. In these terms the   $SU(2)\times SU(2)$ representation is:
\bel
R(u_1, u_2) \vert \alpha; \dot\beta \rangle = (u_1)_{\alpha' \alpha} (u_2)^*_{\dot\beta' , \dot \beta} \vert \alpha' ; \dot \beta' \rangle
\ee
and in terms of $X_{\alpha\dot\beta}$ this is
\bel
X_{\alpha \dot \beta} \rightarrow (u_1 X u_2^{-1})_{\alpha\dot \beta}
\ee

Now we claim there is a unique isomorphism of real vector spaces
\bel
\varphi: \IH \rightarrow V
\ee
such that
\be
\xymatrix{
U(\IH) \times U(\IH) \ar[r]^T \ar[d]^{(\psi, \psi)} & GL(\IH) \ar[d]^{GL(\varphi)} \\
SU(2) \times SU(2) \ar[r]^R  & GL(V) \\
}
\ee
commutes. Here $GL(\varphi)$ is the group isomorphism induced by the isomorphism of real
vector spaces.

Using this distinguished isomorphism we define
\be
\begin{split}
\varphi(1) & = \vert 1 \rangle \\
\varphi(\qi) & = \vert 2 \rangle \\
\varphi(\qj) & = - \vert 3 \rangle \\
\varphi(\qk)  & = - \vert 4 \rangle \\
\end{split}
\ee
The basis $ \{  \vert 1 \rangle, \vert 2 \rangle, \vert 3 \rangle, \vert 4 \rangle \}$
is that given above in equation \eqref{eq:CanonBasis} above. Note the signs in the
last two equations. So, the basis is not {\it that} canonical, but the difference
does not affect the way the Golay code appears.

\end{document}